# Online networks destroy social trust [1]

Fabio Sabatini [2] [*]

Francesco Sarracino [3]

**Abstract**

Studies in the social capital literature have documented two stylised facts: first, a decline in measures of social participation has occurred in many OECD countries. Second, and more recently, the success of social networking sites (SNSs) has resulted in a steep rise in online social participation. Our study adds to this body of research by conducting the first empirical assessment of how online networking affects two economically relevant aspects of social capital, i.e. trust and sociability. We address endogeneity in online networking by exploiting technological characteristics of the pre-existing voice telecommunication infrastructures that exogenously determined the availability of broadband for high-speed Internet. We find that participation in SNSs such as Facebook and Twitter has a positive effect on face-to-face interactions. However, social trust decreases with online interactions. We argue that the rising practice of hate speech may play a crucial role in the destruction of trust.

[1] We gratefully acknowledge the support of the Laboratory for Comparative Social Research at the National Research University Higher School of Economics. We are indebted to Angelo Antoci, Ronald F. Inglehart, Eduard Ponarin, Luca Stanca, Christopher Swader and Christian Welzel for useful comments and suggestions. The paper also benefited from comments by Anna Amalkaeva, Raj Chetty, Hermann Dülmer, Anna Nemirovskaya, Fabio Paglieri, Thomas H. Sander, Marco Simoni, Balazs Telegdy and Bogdan Voicu, by participants at the 3rd LCSR International Workshop (Saint Petersburg, April 26-30, 2013), at the Conference "Cultural and Economic changes under cross-national perspective" (Moscow, November 12-16, 2013), at the Workshop "Social and cultural changes in comparative prospect: values and modernization" (Moscow, March 29-April 6, 2014), and at seminars in Milan and Rome. Usual caveats apply.

[2] Department of Economics and Law, Sapienza University of Rome, Italy, and Laboratory for Comparative Social Research (LCSR), National Research University Higher School of Economics, Moscow and Saint Petersburg, Russia (grant # 11.G34.31.0024 from November 28, 2010).

[*] Corresponding author. Postal address: Sapienza Università di Roma, Facoltà di Economia, via del Castro Laurenziano 9, 00161, Roma, Italy. E-mail: fabio.sabatini@uniroma1.it.

[3] Institut national de la statistique et des études économiques du Grand-Duché du Luxembourg (STATEC), Laboratory for Comparative Social Research (LCSR), National Research University Higher School of Economics, Moscow and Saint Petersburg, Russia (grant # 11.G34.31.0024 from November 28, 2010) and GESIS Leibniz-Institute for the Social Sciences. Email: f.sarracino@gmail.com.

# 1. Introduction

In the years that preceded the social networking revolution, indicators of social participation have declined in many OECD countries (Bartolini et al., 2013; Costa & Kahn, 2003; Putnam, 2002; Sarracino, 2010). However, more recently, the success of social networking sites (SNSs) has resulted in a steep rise in online social participation (Antoci et al. 2013a; 2013b; Brenner and Smith, 2013).

According to the Pew Research Center (PRC) Internet & American Life Project Survey, as of May 2013, 72% of online adults were active on SNSs (67% use Facebook, 16% use Twitter, 15% use Pinterest and 13% use Instagram). Approximately 80% of online young adults (aged 18–29) and 77% of middle-aged adults (30–49) use SNSs (Duggan and Brenner, 2013; Brenner and Smith, 2013). Despite the immensity of these transformations, the impact of online interactions on social capital has so far never been analysed in the economic literature. It is not clear whether in the "social networking era" Internet usage may accelerate the decline in social participation as documented by empirical studies, or if it offers a way to support social relationships against the threats posed by the disruption of ties and the weakening of community life.

A few pioneering economic studies analyse the impact of broadband and, more in general, of Internet use. Pénard and Poussing (2010) find ambiguous results on the relationship between online investments in social capital and the development of face-to-face interactions among Luxembourg Internet users. In a following study, the authors find that non-users are less satisfied with their life than Internet users (Pénard and Poussing 2013). Bauernschuster et al. (2011) show that having broadband Internet at home does not harm social capital in Germany. On the other hand it favours cultural consumption. Campante et al. (2013) show that the diffusion of the broadband may foster specific types of political participation in Italian municipalities. Falck et al. (2012) find that broadband decreases voter turnout in German municipalities. These works are innovative and informative, but they are not able to analyse the role of online networking, which has rapidly become one of the most relevant features of Internet use, or to assess the impact on social trust, which is the most economically relevant aspect of social capital.

Empirical studies in the fields of applied psychology and communication science on the other hand have more specifically analysed how online networking – with a special attention to specific networks such as Facebook and MySpace – influences social interactions across Internet users. However they suffer from the use of survey data collected from strongly biased and non-representative samples, in most cases composed of small communities of undergraduate students.

We innovate this multidisciplinary literature by carrying out the first study on the effect of online interactions through social networking sites (SNSs), chats, newsgroups, and forums, on two economically relevant dimensions of social capital – i.e. generalized trust towards unknown others (hereafter "social trust") and social networks developed through face-to-face interactions among friends and acquaintances – in a large and representative sample of the Italian population. Our main research objective is to investigate whether online networking can support or, by contrast, destroy these two dimensions of social capital.

To reach this goal we use a pooled cross-section of data including the last two waves (2010 and 2011) of the Multipurpose Survey on Households (MHS) provided by the Italian National Institute of Statistics (Istat). This survey contains detailed information on Internet use – with special regard to participation in online networks such as Facebook and Twitter – and the different dimensions of social capital.

Due to the cross-sectional nature of our data we cannot exclude the possibility that online participation may be endogenous to individual social capital. More specifically, there may be three sources of endogeneity: first, it is difficult to distinguish the effect of online networking from that of other phenomena that potentially influence social capital. Second, individual effects, such as personal exogenous shocks, may be correlated with both the propensity for online networking and individual social capital, thus creating a common bias. Third, it is reasonable to suspect the existence of reverse causality: people who meet their friends more frequently, for example, may be encouraged to use online networking to stay in closer touch with them. To deal with these problems we first include in the social capital equations a wide set of individual, household and local level control variables. In addition to usual socio-demographic controls, we place a special focus on the ways in which people connect to the Internet. Then, we instrument participation in SNSs and in chats, newsgroups, and forums, by means of indicators of the local coverage of DSL and optical fibre allowing a fast Internet connection a few years before the collection of MHS data. We illustrate in Section 4.1 how Italy's orography caused a significant variation in access to fast Internet across regions and how such variation is exogenous to social capital and was not driven by individuals' propensity for online networking.

Ordered probit and IV estimates show that participation in SNSs and in chats, newsgroups, and forums is significantly and positively associated with the frequency of meetings with friends and acquaintances. However, we find a significant and negative association between online participation and social trust.

The paper proceeds reviewing the literature on social capital and Internet-mediated interaction. Section IV describes our data and method. Section V specifically addresses endogeneity issues. The empirical results are presented in Sections VI and VII. Section VIII is devoted to the interpretation of results. The conclusion summarizes some lessons on the effects of online networking.

**2. The decline in social capital**

Social capital is generally referred to as all "features of social life – networks, norms, and trust – that enable participants to act together more effectively to pursue shared objectives" (Putnam, 1995: 67). At the level of individuals, Bourdieu (1980), stressed the role of social relations. He argued that actors use relationships as means to increase their ability to advance personal interests. In this context, social capital is "the sum of the resources, actual or virtual, that accrue to an individual or group by virtue of possessing a durable network of more or less institutionalized relationships of mutual acquaintance and recognition" (Bourdieu and Wacquant, 1992: 119, expanded from Bourdieu, 1980: 2). Bourdieu's and Putnam's perspectives describe social capital as a multidimensional concept composed of tangible and intangible features that display their influence both at the micro and the macro level. Uphoff (1999) proposed a classification based on the distinction between structural and cognitive dimensions: in the author's view, structural social capital concerns individuals' behaviours and mainly consists of social participation through various kinds of interpersonal interaction, from informal meetings with friends to active membership in formal organizations. Cognitive social capital derives from individuals' perceptions resulting in trust, values and beliefs that may (or may not) promote pro-social behaviour. In this paper we follow Uphoff's classification to investigate the effect of online networking on the structural and cognitive dimension of social capital as measured by indicators of the frequency of meetings with friends and of social trust[1].

There are several reasons to consider these dimensions of social capital worth investigating in economics. Trust and repeated interaction in networks have been credited with reducing transaction costs, promoting the enforcement of contracts, facilitating credit at the level of individual investors, and to encourage innovation and investment in human and physical capital (see among others Putnam 1993; Fukuyama 1995; Knack and Keefer 1997; Christoforou 2010; Zak and Knack 2011).

Knack (2002) argues that, "Where social mechanisms for the efficient resolution of prisoners' dilemma and principal-agent games are weak or absent (i.e. where most potential pairs of economic transactors cannot trust each other) the private returns to predation increase while the private returns to production fall" (p. 171). Even if these views have been acknowledged in the economic debate only recently, it is worth noting that the concept of the social "embeddedness" of the economic action is deeply rooted in the history of economic thought, and can also be found in the early work of the classical economists. Typical code words of the social capital literature (e.g. trust, altruism, sympathy, and prosocial behaviour) can be found in the work of Adam Smith. In the *Theory of Moral Sentiments*, Smith (1759) argued that there were certain virtues, such as trust and a concern for fairness that, due to their role in the discouragement of cheating, were vital for the functioning of a market economy. Smith described trust as a critical foundation of the early beginnings of the market, allowing the development of trade and economic activities. This point may be reasonably extended by arguing that not only the performance of markets but also, to a larger extent, the resilience of the economic system rely on those institutions (whether formal or informal) that foster the sharing and diffusion of feelings of trust and promote or preserve prosocial behaviour (Andriani and Sabatini, 2013).

Individuals' involvement in networks of relations, on the other hand, has been found to be significantly and positively correlated with happiness (Bruni and Stanca, 2008; Bartolini and Bilancini, 2010; Bartolini et al., 2013), self-esteem (Ellison et al., 2007; Steinfield et al., 2008), physical and mental health (Rocco et al., 2011; Yamamura, 2011a), income (Robison et al., 2011), and entrepreneurship (Bauernschuster et al., 2010). Social isolation has been found to be a strong predictor of bad health conditions and poor levels of well-being (Kawachi et al., 2011; Yamamura, 2011b).

How are these dimensions of social capital performing in recent years? In his best-seller *Bowling Alone*, Robert Putnam (2000) draws on various sources to document that a decline in social participation measures – such as membership in formal organizations, the intensity of members' participation, informal social connectedness, and interpersonal trust – began in the United States in the 1960s and 1970s with a sharp acceleration in the 1980s and 1990s.

The "decline of community life thesis" (Paxton, 1999, p. 88) advanced by Putnam prompted a number of subsequent empirical tests. Based on the General Social Surveys (GSS) data for the period 1975–94, Paxton (1999) finds some decline in the general measure of social capital (given by a combination of trust and membership in associations), a decline in interpersonal trust, and no decline in associations. Costa and Kahn (2003) use a number of different sources to assess the development of social capital in the United States since 1952 by evaluating trends in participation and community life. The authors find a decline in indicators of volunteering, membership in organizations, and entertainment with friends and relatives. Bartolini et al. (2013) use GSS data to investigate the evolution of social connections – measured through membership in Putnam and Olson groups[2] and indicators of perceived trustworthiness, helpfulness and fairness, and confidence in institutions in the United States between 1975 and 2002, and find that it generally shows a declining trend. Bartolini and Bonatti (2008) explain how this negative trend may be reconciled with the satisfactory growth performance of the U.S. through a theoretical framework modelling the hypotheses that the expansion of market activities weakens social capital formation, and that firms utilize more market services in response to the declining social capital.

Apart from the United States, there seems to be a common pattern of declining trust, political participation, and organizational activity across industrialised democracies during the 1980s and 1990s, with the exception of China, Japan, Korea and the Scandinavian countries (Chen & Gao, 2013; Lee, 2008; Leigh, 2003; Listhaug & Grønflaten, 2007). Declining trends of one or more dimensions of social capital have been documented for England and Wales over the period 1972–1999 (Li et al., 2003), Great Britain over 1980–2000 (Sarracino, 2010), and Australia over 1960–1990 (Cox, 2002)[3].

## 3. The role of Internet-mediated interaction

Putnam (2000) discusses three main explanations for the decline in American social capital: 1) the reduction in the time available for social interaction – related to the need to work more, to the rise in labour flexibility, and to the increase in commuting time in urban areas; 2) the rise in mobility of workers and students; and 3) technology and mass media.

In the last decade, Putnam's arguments have found support in a number of studies investigating the effect exerted on various dimensions of social connectedness by the rise in working time (Bartolini & Bilancini, 2011), labour mobility (Routledge & von Ambsberg, 2003), urban sprawl and commuting (Besser et al., 2008; Wellman, 2001)[4], and by the social poverty of the surrounding environment, which can prompt individuals to pursue social isolation (Bartolini and Bonatti, 2003; Antoci, Sacco and Vanin, 2007; Antoci, Sabatini and Sodini, 2012; 2013a; 2013b).

Putnam's argument about the role of technology and media in the evolution of social interaction, on the other hand, is widely debated in the literature. The author's explanation of the possibly negative role of technology was centred on the socially detrimental effects of television and other forms of "private" entertainment such as video games. This concern was shared by the early sociological literature on Internet use, which basically developed two main arguments. First, the more time people use the Internet for leisure, the more time has to be detracted from social activities like communicating with friends, neighbours, and family members (Nie, 2001; Nie et al., 2002; Gershuny, 2003; Wellman et al., 2001). This argument was proposed by studies that date back to shortly before the explosion of online networking and it did not differentiate between pure entertainment and social activities. At that time, using the Internet was predominantly a solitary pastime like watching TV or reading newspapers.

A second argument relies on the concept of "community without propinquity" (Webber, 1963) and on the earlier theories of the Chicago School of Sociology. In a famous paper, Wirth (1938) claimed that any increase in the heterogeneity of the urban environment would have provoked the cooling of "intimate personal acquaintanceship" and would result in the "segmentation of human relations" into those that were "largely anonymous, superficial, and transitory" (Wirth, 1938, p. 1). This argument can be easily applied to the Internet, which seems to have the potential to fragment local communities into new virtual realities of shared interest that may negate the necessity of face-to-face encounters (Antoci et al., 2012). The "anonymization hypothesis", however, has been challenged by results from studies specifically targeted at verifying the effects of online networking on communities living in a precise and limited geographic location, such as a city area or suburb.

In one of the rare studies on online networking that were conducted in the 90s, Hampton and Wellman (2003) drew on survey and ethnographic data from a wired suburb of Toronto and found that high-speed always-on access to the Internet, coupled with a local online discussion group, transformed and enhanced relationships among neighbours. In the Toronto sample, Internet use strengthened weaker social ties without causing any deterioration in the already steady relationships. In the authors' words,

"not only did the Internet support neighbouring, it also facilitated discussion and mobilization around local issues" (Hampton and Wellman 2003, p. 277).

Sceptical findings about the relational effects of Internet use have not found support in more recent empirical studies conducted in applied psychology and communication science after the "explosion" of online networks. All the studies mentioned above exclusively refer to face-to-face interactions and completely disregard online participation. However in the past few years, Internet-mediated interaction has literally revolutionised individuals' social lives. In contrast to the early age of the Internet, today the use of the Internet is strongly related to being connected to SNSs, which in turn entails engagement in social activities.

According to a survey conducted by Princeton Survey Research Associates International in November 2010 among a sample of 2,255 adults, SNSs are used increasingly to keep up with close social ties; the average user of an SNS has closer ties and is half as likely to be socially isolated as the average American; and finally, Facebook users are more trusting, have closer relationships and are much more politically engaged than the average American. Internet users can gather more support from their social ties than those who do not use the Internet. Facebook users get the most support; it has been found that Facebook plays a crucial role in reviving "dormant" relationships (Brenner, 2013; Hampton et al., 2011). More than half the Internet users create and share original content online. According to a nationally representative survey of 1,000 adults conducted in October 2013, 54% of adult users post original photos or videos online that they themselves have created (Duggan, 2013). Sharing photos is a fundamental way to keep relatives, friends, and acquaintances posted on personal experiences, a method which has proven to be particularly effective for people such as workers and students living away from home. Overall, 39% of all American adults took part in some sort of political activity on an SNS during the 2012 campaign. In 2012, 17% of all adults posted links to political stories or articles on SNSs, and 19% posted other types of political content. In 2012, 12% of all adults followed or friended a political candidate or other political figure on an SNS, and 12% belonged to an SNS group of a socio-political nature (Smith, 2013). In December 2010, U.S. Internet users were found to be more likely than others to be active in some kind of voluntary group or organization: 80% of American Internet users participated in groups, compared to 56% of non-Internet users. Moreover, social media users are even more likely to be active: 82% of social network users and 85% of Twitter users are group participants (Rainie et al., 2011).

These figures mark a dramatic increase from February 2005, when PRC began to monitor Internet usage in the U.S. (Madden & Zickuhr, 2011), and begs reconsideration of the alleged social isolation that we commonly associate with intense Internet usage.

Findings from recent empirical studies support the hypothesis that online interactions may play a positive role in the preservation and development of social ties against the threats posed by the weakening of community life and the gradual erosion of social capital. Authors have claimed that SNSs support the strengthening of bonding and bridging social capital (Lee, 2013; Steinfield et al., 2008), children's social activities (Bauernschuster et al., 2011) as well as the social integration and well-being of the elderly (Näsi et al., 2012; Russel et al., 2008). SNSs may allow the crystallization of weak or latent ties which might otherwise remain ephemeral (Ellison et al., 2007; Haythornthwaite 2005), help users cope with social anxiety and bouts of negativity and loneliness (Clayton et al., 2013; Grieve et al., 2013; Morahan-Martin and Schumaker, 2003), boost teenagers' self-esteem by encouraging them to relate to their peers (Cheung et al., 2010; Ellison et al., 2011; Trepte and Reinecke, 2013), promote civic engagement and political participation (Gil de Zuniga, 2012; Kittilson & Dalton, 2011; Gil de Zuniga 2012; Zhang et al., 2010), stimulate social learning and improve cognitive skills (Alloway et al., 2013; Burke et al., 2011), enhance social trust (Valenzuela et al., 2009) and help the promotion of collective actions (Chu and Tang, 2005)[5].

Drawing on survey data from a random sample of 800 undergraduate students, Ellison et al. (2007) find that certain types of Facebook use can help individuals accumulate and maintain social capital. Their results support the hypothesis that social networks help students to overcome the barriers to participation so that individuals who might otherwise shy away from initiating communication with others are encouraged to do so through something like the Facebook infrastructure. In the authors' words, highly engaged users are using Facebook to "crystallize" relationships that might otherwise remain ephemeral.

Steinfield et al. (2008) analysed panel data from two surveys on Facebook users conducted a year apart at a large U.S. university. Intensity of Facebook use in year one strongly predicted bridging social capital outcomes in year two (even after controlling for measures of self-esteem and satisfaction with life). The authors suggest that interactions through Facebook "help reduce barriers that students with lower self-esteem might experience in forming the kinds of large, heterogeneous networks that are sources of bridging social capital" (Steinfield et al., 2008, pp. 434). However, the literature on Facebook suggests that social networks – and, more generally, Internet-mediated communication – serves more to preserve relations among offline contacts than to activate latent ties or create connections with strangers (Ellison et al., 2007). In the field of economics, a recent paper based on data drawn from the 2008 section of the German Socio-Economic Panel and confidential data provided by *Deutsche Telekom*, Bauernschuster et al. (2011) find that having broadband Internet access at home has positive effects on an individual, manifesting in his frequency of visiting theatres, opera and exhibitions, and in his frequency of visiting friends. The authors address endogeneity issues by instrumenting broadband access through the availability of appropriate infrastructures, which was in turn related to an unforeseeable "technological accident" which exogenously jeopardized individuals' access to broadband. Exploring a sub-sample of children aged 7 to 16 living in the sampled households, the authors further found evidence that having broadband Internet access at home increases the number of children's out-of-school social activities such as learning sports, ballet, music, painting, or joining youth clubs.

Using data on Italian municipalities, Campante et al. (2013) find that the impact of broadband availability on political participation "changes over time and is crucially affected by the reaction of the political supply side" (p. 3). The authors show that the diffusion of broadband led, initially, to a significant decline in electoral turnout in national parliamentary elections between 1996-2001 (pre-broadband) and 2006-2008 (post-broadband). This initial negative effect of Internet on turnout was largely reversed in the following elections, held in 2013. Falck et al. (2012) conduct a similar analysis drawing on data on German municipalities. The authors find that an increase in DSL availability significantly decreases voter turnout. Analysing German municipality-level data for the period 2002-2005, Czernich (2012) obtains the opposite result that Internet broadband fosters electoral participation.

These studies add to the literature by addressing the role of broadband Internet on forms of participation at the individual and local level. However, due to lack of data, their authors could not tackle the effect of online networking, nor at the individual neither at the aggregated level.. We further add to this body of research providing the first attempt to assess the effect of online networking – in the form of participation to social networks such as Facebook and Twitter, and in forums, chat rooms, and newsgroups – on trust and sociability in a large and representative sample of the Italian population.

**4. Data and methods**

We use a pooled cross-section of data drawn from the last two waves (2010 and 2011) of the Multipurpose Survey on Households (MHS) provided by the Italian National Institute of Statistics (Istat). This survey investigates a wide range of social behaviours and perceptions by means of face-to-face interviews on a nationally and regionally representative sample of approximately 24,000 households, roughly corresponding to 50,000 individuals.

As mentioned in the Introduction, we measure social capital through indicators of its structural and cognitive dimension. The structural dimension is given by social interactions ($social\_interactions_i$), as measured by the frequency of meetings with friends. Respondents were asked to report how many times they meet their friends on a scale from 1 (in case they have no friends) to 7 (if respondents meet their friends everyday)[6]. Cognitive social capital is given by social trust ($trust_i$), as measured by binary responses to the question: "Do you think that most people can be trusted, or that you can't be too careful in dealing with people?" as developed by Rosenberg (1956).

In addition, we also employ as dependent variable a further indicator of social trust drawn from the so-called "wallet question": "Imagine you lost your wallet with your money, identification or address in your city/area and it was found by someone else. How likely do you think your wallet would be returned to you if it were found by a neighbour/the police/a stranger?" Possible responses were: "Very likely", "Fairly likely", "Not much likely", and "Not likely at all". The introduction of wallet questions into surveys was spurred by experiments reported in Reader's Digest Europe in April 1996 (and subsequently discussed in the Economist, June 22, 1996). These experiments involved dropping 10 cash-bearing wallets in each of 20 cities in 14 western European countries, and in each of a dozen US cities (Helliwell and Wang, 2011). The data on the frequency of wallet returns were later used by Knack (2001) to provide some behavioural validation for the use of answers to the "Rosenberg question" on generalized trust. Knack (2001) found that at the national level the actual frequency of the returns correlated at the 0.65 ($p < 0.01$) level with national average responses to the general social or interpersonal trust question (as measured by the World Values Survey). While this provides strong validation for the meaningfulness of international differences in survey responses to social trust questions, it also suggests a way of adding more specific trust questions to surveys. Here we followed Knack (2001) and measured social trust based on the responses to the hypothesis that the wallet was found by a complete stranger. We reversed the scale, so that larger values indicate greater trust in unknown others.

Online networking is given by two dichotomous variables capturing respondent *i*'s participation in social networking sites such as Facebook, Twitter, and MySpace and in chats, forums, and newsgroups. The relationship between the two categorical indicators of social capital (the frequency of meetings with friends and responses to the wallet question) and online networking was investigated through an ordered probit model with robust standard errors reporting marginal effects. If the dependent variable is ordered in *K* categories, then the model for social interactions is:

$$y_i = \begin{cases} 1 & \text{if } \quad y_i \leq 0 \\ 2 & \text{if } \quad 0 < y_i \leq c_1 \\ 3 & \text{if } \quad c_1 < y_i \leq c_2 \\ \cdot \\ \cdot \\ \cdot \\ K & \text{if } \quad c_{K-1} < y_i \end{cases} \quad (1)$$

where $0 < c_1 < c_2 < ... < c_{K-1}$; $y_i = \alpha + \beta_1 \cdot fb_i + \beta_2 \cdot chat_i + \theta \cdot X_i + \varepsilon_i$, $\varepsilon_i \sim N(0,1)$. $c_{K-1}$ are unknown parameters to be estimated, and $\theta$ is a vector of parameters for the vector of control variables $X_i$. To explore the relationship between the dichotomous measure of social trust and online networking we employed a probit model with robust standard errors reporting marginal effects. For individual *i*, the trust equation is:

$$social\_trust_i = \begin{cases} 1 & \text{if } \quad y_i > 0 \\ 0 & \text{if } \quad y_i < 0 \end{cases} \quad (2)$$

where $y_i = \alpha + \beta_1 \cdot fb_i + \beta_2 \cdot chat_i + \theta \cdot X_i + \varepsilon_i$, $\varepsilon_i \sim N(0,1)$

The list of control variables includes:

- the kind of technology that respondents used to connect to the Internet. Possible categories were cable broadband (optical fibre, intranet, PLC, etc.), satellite or other wireless connections (e.g. wi-fi and wi-max), wireless connection through tablets and/or mobile phones employing a 3G mobile telecommunication technology, wireless connection employing a 3G modem (e.g. a USB key), or connection with a WAP or a GPRS mobile phone.

- Age (both in linear and squared form), gender, marital status, number of children, education, work status[7], and the time spent in commuting (in minutes).

We accounted for commuting for two main reasons. First, the time spent on commuting may be distracted from social interactions. Second, it may be considered as a proxy for spatial fragmentation which allows us to test one of Putnam's claims on the detrimental effects of the spread of modern cities. In the author's words: "It is not simply time spent in the car itself, but also spatial fragmentation between home and workplace, that is bad for community life" (Putnam 2000, pp. 213-214).

A summary of descriptive statistics is presented in Table 1.

| Table 1. Descriptive statistics | | | | | |
|---|---|---|---|---|---|
| Variables | Obs | Mean | St. dev. | Min | Max |
| Frequency of meetings with friends | 78988 | - | - | 1 | 7 |
| Social trust (Rosenberg question) | 77723 | 0.223 | 0.416 | 0 | 1 |
| Social trust (wallet question) | 77368 | 1.623 | 0.726 | 1 | 4 |
| Use of SNSs | 35282 | 0.453 | 0.498 | 0 | 1 |
| Use of chats, forums, newsgroups | 17270 | 0.351 | 0.477 | 0 | 1 |
| Woman | 79433 | 0.521 | 0.500 | 0 | 1 |
| Age | 79433 | 50.11 | 18.21 | 18 | 90 |
| Age squared | 36111 | 28.43 | 19.07 | 3.240 | 81 |
| Minutes spent on commuting | 79433 | 18.67 | 12.32 | 0 | 57 |
| Civil status | 79433 | - | - | 1 | 4 |
| Educational qualification | 79433 | - | - | 1 | 5 |
| Work status | 79433 | - | - | 1 | 7 |
| Number of children | 79433 | 1.011 | 1.009 | 0 | 7 |
| Frequency of meeting friends (by region) | 79433 | 5.104 | 0.168 | 4.87 | 5.51 |
| High education (% by region) | 79433 | 11.30 | 1.614 | 7.80 | 16.02 |
| Real GDP per capita (thousands of euro 2005) | 79433 | 22.92 | 5.746 | 14.88 | 30.77 |
| Region | 79433 | - | - | 10 | 200 |
| Year | 79433 | - | - | 2010 | 2011 |

## 5. Endogeneity issues

The coefficients from equations (1) and (2) indicate the sign and magnitude of partial correlations among variables. However, we cannot discard the hypothesis that online networking is endogenous to social interactions and social trust. Individual effects such as personal characteristics or exogenous shocks may be correlated with both online networking and the two dimensions of social capital we account for. Outgoing and open-minded persons who have a higher propensity for trusting strangers may also be more attracted by new forms of socialization such as Facebook or chats. Individuals responding yes to the question "if most people can be trusted" may have a higher propensity for developing new social ties and may be more attracted to new forms of socialization like Facebook and chats. Or, for example, they may be more willing to seek help from strangers in forums and newsgroups in case of troubles with computers or other electronic devices. By contrast, individuals who trust strangers less may find chats and newsgroups unattractive. As illustrated in sections I and IV, we tried to reduce the possible influence of omitted variables through the introduction of a large set of covariates in our models.

However, and most importantly, reverse causality might also arise. For example, people who meet their friends frequently may be encouraged to join online networks to strengthen existing social ties. Reverse causality may also work in the opposite direction to the extent to which people who have no (or just a few) friends may look for interactions on Facebook to alleviate their social isolation.

To deal with these problems, we turn to instrumental variables estimates using a two stage least squares (2SLS) model (Wooldridge, 2002) where, in the first stage, we instrument our two measures of online networking.

A reliable instrumental variable must meet at least two criteria. First, it must be theoretically justified and statistically correlated with online networking ("relevance" condition), after controlling for all other exogenous regressors. Second, it must be uncorrelated with the disturbance term of the two social capital equations ("orthogonality" condition).

We identified two econometrically convenient instruments in: 1) the percentage of the population for whom a DSL connection was available in respondents' region of residence according to data provided by the Italian Ministry of Economic Development. DSL (digital subscriber line, originally digital subscriber loop) is a family of technologies that provides Internet access by transmitting digital data over the wires of a local telephone network. Basically, it is a way to improve the speed of data transmission through old telephonic infrastructures. 2) A measure of the digital divide given by the percentage of the region's area that was not covered by optical fibre, elaborated from data provided by The Italian Observatory on Broadband. Optical fibre permits transmission over longer distances and at higher bandwidths (data rates) than other forms of communication.

Both the instrumental variables were measured in 2008, two years before the first wave of the Multipurpose Household Survey, which we employ in our study.

We believe that the 2008 level of regional DSL coverage cannot *per se* exert a direct influence on individual social capital. Rather, the availability of DSL infrastructures in the area creates the premise for the individual choice to purchase a fast-speed access and, subsequently, to develop online interactions through social networking sites, chats, forums and newsgroups. In other words, it is reasonable to assume that the impact of broadband coverage on social capital solely occurs through the use of social networking sites, chats, forums, newsgroups and similar forms of Internet-mediated communication.

It is easy to demonstrate that the variation in the availability of DSL is exogenous to social capital. DSL technology relies on the transmission of data over the user's copper telephone line, i.e. over pre-existing voice telecommunications infrastructures. However, the existence of a telephone infrastructure is just a necessary and not sufficient condition for the availability of the broadband. What matters is in fact the so-called "local loop", i.e. the distance between final users' telephone line and the closest telecommunication exchange or "central office" (Falck et al., 2012; Czernich, 2012; Campante et al., 2013).

For the provision of traditional voice services, the length of this distance does not influence the quality of the connection. However, for the provision of DSL this distance matters since the longer the copper wire, the less bandwidth is feasible via this wire. If the distance is above a certain threshold (approximately 4.2 kilometres, corresponding to 2.61 miles), then the band of the copper wires serving telephone communications cannot be wide enough to support a fast Internet connection (Falck et al.

2012; Czernich, 2012). It is thus impossible to implement the broadband connection through traditional copper wires. This is the case of Italian rural areas, which constitute more than half of the Italian territory and are mostly composed of severely isolated and low densely populated highlands or hills. In 2007 these areas were generally characterised by the high length ($\geq$ 4.2 kilometres) of local loops, which ultimately depended on the imperviousness of the territory. As a consequence, these areas in most cases lacked the infrastructures needed for the diffusion of the DSL broadband (Ciapanna and Sabbatini, 2008; Agcom, 2011).

The distribution of such infrastructures in 2008 should thus be considered as exogenous to the level of social capital in 2010-11 because it strictly depended on local loops, whose location was determined several decades before the advent of the Internet, based on the orographic features of the territory (Agcom, 2011; Campante et al., 2013). In Figure 4 in the Appendix, we report a comparison between a map illustrating the orographic characteristics of the Italian territory and a map showing broadband coverage in 2007 helps to better understand the extent to which broadband diffusion was determined by exogenous, orographic, factors.

To further check the validity of this instrument, we reviewed the literature and found that DSL coverage in the region of residence has never been found to be correlated with social interactions and social trust at the individual level. The study of Bauernschuster et al. (2011) investigated the role of *individuals'* use of broadband on social interactions and cultural consumption. The availability of appropriate technological infrastructures in the area of residence was used by the authors to instrument the individual choice to purchase a broadband access for connecting to the Internet. Similar instruments were used in municipality-level studies on electoral participation by Falck et al. (2012), Czernich (2012), and Campante et al. (2013). In Bauernschuster et al. (2011), broadband access was then shown to positively affect social interactions. This result is supported by our estimates which, thanks to the wealth of our dataset, allow us to further improve our understanding of the role of the Internet by showing which kind of use may specifically affect social capital (see Sections VI and VII). On the other hand, the DSL coverage in 2008 cannot of course be endogenous – in the sense of reverse causality – to the individual involvement in online networks in 2010-2011. The possibility of common bias between the two variables also seems unlikely. One could argue that individuals who exhibited a positive propensity for participation in SNSs in the 2010-2011 period may have had a higher propensity for promoting actions aimed at extending the regional broadband coverage in 2008. However, as we explained above, the reasons for the digital divide across Italian regions are basically exogenous and linked to the orographic features of the territory. In addition, it must be noted that in Italy, Facebook, Twitter and other social networking sites only boomed after 2008[8].

The arguments supporting the assumption of the orthogonality of the share of the population covered by DSL are even stronger for the second instrument. When, as we explained above, the broadband connection cannot be implemented through pre-existing copper wires, it is necessary to turn to an optical fibre-based technology. The possibility and the costs of installing this type of infrastructures, however, even more strongly rely on the exogenous characteristics of the natural environment. Differently from DSL, in fact, optical fibre entails the need to install new cables underground.

This involves excavation works, which are very costly and generally delay or even prohibit the provision of broadband in the area. Once again, orographic differences between regions must be considered as a "natural" cause of the digital divide which generated a variation in access to fast Internet across regions that is exogenous to people's social capital and cannot be driven by their preference for online networking.

The assumption of orthogonality of the instruments is not disconfirmed by the tests of over-identifying restrictions we run in the context of IV estimates (reported in Section VI).

For any given set of orographic characteristics of the area, the provision of the broadband – whether through the DSL or the optical fibre technology – may also have been influenced by some socio-demographic factors that affected the expected commercial return of the provider's investment, such as population density, per capita income, the median level of education and the local endowments of social capital. These characteristics could be expected to correlate with our outcomes of interest in ways that could confound causal interpretation. To account for the confounding effect of these characteristics, we included a set of regional level controls in our regressions

Regarding the relevance of the instruments, the discussion about how the digital divide may influence SNSs is not trivial. There are in fact two ways in which the digital divide can influence individuals' propensity for online networking. On the one hand, it can be argued that the bigger the area covered by cable infrastructures, the higher should be the individual propensity for online networking. However, in areas where broadband access is less diffuse, the use of social networking sites is a scarce commodity. In these places the demand for broadband may be higher as consumers are keen to participate in SNSs with any available device. If this is the case, the individual propensity for networking should be positively correlated with the scarcity of the broadband.

The relevance of instruments will be further discussed in Section VII (presenting results of IV estimates) as it is strictly related to evidence from the first step of IV regressions.

## 6. Results

Table 2 presents estimates of equation (1). In model 1 we report correlations of the dependent variable with covariates we controlled for. Face to face interactions are found to be significantly and negatively correlated with age and with the amount of time spent in commuting. Women also meet their friends less frequently. In model 2 we introduce participation in social networking sites, which is found to be significantly and positively correlated with face-to-face interactions. Model 3 highlights a significant and positive association between face-to-face interactions and participation in chats, forums, and newsgroups. In model 4 we simultaneously account for the two forms of networking, which are confirmed to be significantly and positively associated with the frequency of encounters with friends and acquaintances. In model 5 we include some aggregate controls at the region level. All models have also been estimated in a reduced sample (*n* = 10745) only including the observations in Model 5 (estimates are not reported for the sake of brevity and are available upon request).

Table 3 presents estimates of equation (2) on social trust. Women exhibit significantly lower levels of trust, which is also shown to be U-shaped with age. Networking via SNSs is significantly and positively associated with social trust. However, when we also account for participation in chats, newsgroups, and forums in the structural equation the coefficient of SNSs loses its statistical significance.

In Table 4 we report estimates of equation (1) where responses from the "wallet question" are used to proxy a further indicator of social trust. The two measures of online networking are found to be significantly and negatively correlated with trust in strangers. However, if we jointly account for both the indicators of networking in the same regression, their correlation with the dependent variables loses its statistical significance. Social trust is also U-shaped with age.

Table 2. Online networking and face to face interactions: ordered probit estimates

|  | Model 1 | Model 2 | Model 3 | Model 4 | Model 5 |
|---|---|---|---|---|---|
| Dependent variable: Frequency of meetings with friends | | | | | |
| *Type of connection to the Internet* | | | | | |
| Dsl (d) | 0.0276 (1.17) | 0.0167 (0.62) | 0.0637 (1.63) | 0.0541 (1.38) | 0.050 (1.29) |
| Fibre (d) | -0.0343 (-0.61) | -0.0504 (-0.81) | -0.0130 (-0.15) | -0.0237 (-0.27) | -0.026 (-0.30) |
| Satellite (d) | 0.0553* (1.77) | 0.0255 (0.73) | 0.0726 (1.42) | 0.0614 (1.20) | 0.056 (1.10) |
| 3G (d) | 0.0275 (0.57) | -0.00911 (-0.17) | -0.0585 (-0.74) | -0.0675 (-0.86) | -0.072 (-0.92) |
| USB (d) | -0.00352 (-0.13) | 0.00663 (0.22) | 0.0168 (0.36) | 0.00507 (0.11) | 0.003 (0.08) |
| Mobile (d) | -0.0456 (-0.92) | -0.0933* (-1.67) | -0.0719 (-1.13) | -0.0843 (-1.32) | -0.091 (-1.44) |
| *Main demographic, social and economic characteristics* | | | | | |
| Women (d) | -0.194*** (-14.66) | -0.183*** (-12.42) | -0.189*** (-8.92) | -0.187*** (-8.82) | -0.186*** (-8.76) |
| Age | -0.0884*** (-18.56) | -0.0896*** (-16.85) | -0.0993*** (-13.25) | -0.0976*** (-13.00) | -0.097*** (-12.98) |
| Age squared / 100 | 0.0851*** (15.39) | 0.0890*** (14.21) | 0.0965*** (10.94) | 0.0954*** (10.80) | 0.095*** (10.77) |
| Minutes spent on commuting | -0.00210*** (-4.00) | -0.00227*** (-3.92) | -0.00179** (-2.16) | -0.00174** (-2.11) | -0.002** (-2.26) |
| *Regional controls* | | | | | |
| Frequency of meeting friends (by region) | | | | | 0.815*** (7.60) |
| High education (% by region) | | | | | -0.005 (-0.76) |
| real GDP per capital | | | | | -0.007** (-2.23) |
| *Indicators of online networking* | | | | | |
| Participation in social networking sites (d) | | 0.163*** (10.13) | | 0.132*** (4.87) | 0.132*** (4.88) |
| Participation in chats, forums and newsgroups (d) | | | 0.148*** (6.29) | 0.0788** (2.84) | 0.075** (2.72) |
| Observations | 27068 | 22148 | 10765 | 10745 | 10745 |
| Pseudo R$^2$ | 0.084 | 0.096 | 0.100 | 0.101 | 0.100 |

Regressions include socio-demographic and year controls: variables are omitted for the sake of brevity and are available upon request to the authors.
d = for discrete change of dummy variable from 0 to 1
* $p < 0.1$, ** $p < 0.05$, *** $p < 0.001$

Table 3. Online networking and social trust: probit estimates

| | Model 1 | Model 2 | Model 3 | Model 4 | Model 5 |
|---|---|---|---|---|---|
| Dependent variable: social trust | | | | | |
| *Type of connection to the Internet* | | | | | |
| Dsl (d) | -0.0134 | 0.0325 | -0.0602* | 0.00218 | 0.003 |
| | (-0.53) | (0.97) | (-1.65) | (0.05) | (0.08) |
| Fibre (d) | -0.00210 | 0.0328 | -0.0301 | -0.0332 | -0.020 |
| | (-0.04) | (0.42) | (-0.37) | (-0.30) | (-0.19) |
| Satellite (d) | 0.0516 | 0.106 | 0.0436 | 0.184** | 0.184** |
| | (1.54) | (2.44) | (0.92) | (2.91) | (2.92) |
| 3G (d) | 0.00158 | 0.0500 | -0.0146 | 0.132 | 0.156 |
| | (0.03) | (0.76) | (-0.22) | (1.30) | (1.54) |
| USB (d) | -0.00184 | 0.0598 | -0.0410 | 0.0616 | 0.069 |
| | (-0.06) | (1.56) | (-0.99) | (1.09) | (1.22) |
| Mobile (d) | 0.0404 | 0.0763 | 0.0164 | 0.0918 | 0.091 |
| | (0.69) | (1.11) | (0.23) | (1.17) | (1.18) |
| *Main demographic, social and economic characteristics* | | | | | |
| Women (d) | -0.0865*** | -0.0821*** | -0.0801*** | -0.0701** | -0.068** |
| | (-6.11) | (-4.34) | (-4.12) | (-2.58) | (-2.51) |
| Age | 0.0151** | 0.0152** | 0.0162** | 0.0250** | 0.025** |
| | (3.21) | (2.39) | (2.46) | (2.71) | (2.71) |
| Age squared / 100 | -0.00398 | 0.00227 | -0.00165 | -0.00664 | -0.006 |
| | (-0.74) | (0.31) | (-0.22) | (-0.62) | (-0.63) |
| Minutes spent on commuting | -0.000404 | -0.00106 | 0.000078 | -0.000397 | -0.0001 |
| | (-0.72) | (-1.44) | (0.10) | (-0.37) | (-0.56) |
| *Regional controls* | | | | | |
| Frequency of meeting friends (by region) | | | | | 0.178 |
| | | | | | (1.32) |
| High education (% by region) | | | | | -0.056*** |
| | | | | | (-5.93) |
| real GDP per capial | | | | | 0.038*** |
| | | | | | (9.15) |
| *Indicators of online networking* | | | | | |
| Participation in social networking sites (d) | | 0.0587** | | 0.0256 | 0.024 |
| | | (2.87) | | (0.74) | (0.70) |
| Participation in chats, forums and newsgroups (d) | | | 0.0345 | 0.0653* | 0.062* |
| | | | (1.35) | (1.85) | (-2.04) |
| Observations | 39960 | 22074 | 20944 | 10720 | 10720 |
| Pseudo $R^2$ | 0.038 | 0.040 | 0.040 | 0.044 | 0.042 |

Regressions include socio-demographic and year controls: variables are omitted for the sake of brevity and are available upon request to the authors.
d = for discrete change of dummy variable from 0 to 1
* $p < 0.1$, ** $p < 0.05$, *** $p < 0.001$

Table 4. Online networking and social trust measured through the "wallet question": ordered probit estimates

| | Model 1 | Model 2 | Model 3 | Model 4 | Model 5 |
|---|---|---|---|---|---|
| Dependent variable: social trust measured through the "wallet question" | | | | | |
| *Type of connection to the Internet* | | | | | |
| Dsl (d) | 0.0105 | 0.0192 | -0.000401 | 0.0368 | 0.041 |
|  | (0.50) | (0.69) | (-0.01) | (0.94) | (1.06) |
| Fibre (d) | 0.0693 | 0.109* | 0.0696 | 0.109 | 0.108 |
|  | (1.49) | (1.80) | (1.09) | (1.25) | (1.24) |
| Satellite (d) | 0.0305 | 0.0563 | 0.0239 | 0.0584 | 0.054 |
|  | (1.08) | (1.52) | (0.60) | (1.09) | (1.01) |
| 3G (d) | 0.0401 | 0.0866 | 0.00640 | 0.150* | 0.162* |
|  | (1.03) | (1.58) | (0.12) | (1.77) | (1.93) |
| USB (d) | -0.0245 | -0.0139 | -0.0333 | -0.00460 | 0.008 |
|  | (-1.03) | (-0.44) | (-0.98) | (-0.10) | (0.18) |
| Mobile (d) | 0.0102 | 0.0194 | -0.00115 | 0.0289 | 0.026 |
|  | (0.21) | (0.33) | (-0.02) | (0.44) | (0.40) |
| *Main demographic, social and economic characteristics* | | | | | |
| Women (d) | -0.00970 | 0.0127 | 0.00933 | 0.0457** | 0.047** |
|  | (-0.83) | (0.80) | (0.58) | (2.00) | (2.11) |
| Age | 0.0263*** | 0.0271*** | 0.0285*** | 0.0350*** | 0.034*** |
|  | (6.71) | (5.08) | (5.20) | (4.49) | (4.39) |
| Age squared / 100 | -0.0213*** | -0.0182** | -0.0222*** | -0.0260** | -0.025** |
|  | (-4.73) | (-2.92) | (-3.49) | (-2.86) | (-2.82) |
| Minutes spent on commuting | -0.000487 | -0.000474 | -0.000771 | -0.000691 | -0.001 |
|  | (-1.04) | (-0.77) | (-1.19) | (-0.78) | (-1.32) |
| *Regional controls* | | | | | |
| Frequency of meeting friends (by region) | | | | | -0.136 |
|  | | | | | (-1.22) |
| High education (% by region) | | | | | -0.077*** |
|  | | | | | (-9.55) |
| real GDP per capital | | | | | 0.035*** |
|  | | | | | (10.46) |
| *Indicators of online networking* | | | | | |
| Participation in social networking sites (d) | | -0.0425** | | -0.0210 | -0.025 |
|  | | (-2.50) | | (-0.72) | (-0.87) |
| Participation in chats, forums and newsgroups (d) | | | -0.0480** | -0.00196 | -0.008 |
|  | | | (-2.24) | (-0.07) | (-0.27) |
| Observations | 39901 | 22081 | 20922 | 10711 | 10711 |
| Pseudo $R^2$ | 0.025 | 0.027 | 0.025 | 0.030 | 0.026 |

Regressions include socio-demographic and year controls: variables are omitted for the sake of brevity and are available upon request to the authors.
d = for discrete change of dummy variable from 0 to 1
* $p < 0.1$, ** $p < 0.05$, *** $p < 0.001$

Our instrumental variables approach uses the percentage of the population for whom DSL connection was available in respondents' area of residence in 2008 and the percentage of the region's area that was not covered by optical fibre in 2008 as instruments for the individual propensity for online networking in the period 2010-2011. Our two-stage model can be described by the following two equations:

$$online\_networking_i = \pi_1 + \pi_2 \cdot dsl + \pi_3 \cdot fiber + \pi_4 \cdot W_i + v_i \quad (4)$$

$$social\_capital_i = \alpha + \theta \cdot X_i + \gamma_1 \cdot dsl + \gamma_2 \cdot fiber + \mu_i \quad (5)$$

To assess the effect of online networking on face-to-face interactions, equation (4) is estimated using a probit model and equation (5) is estimated using an ordered probit model[9]. Estimated coefficients are reported in Table 5.

The relationship between online networking and social trust, as measured through responses to the Rosenberg question, is then estimated using a probit model in both the stages of the procedure. Results are reported in Table 6.

When we use the alternative measure of social trust obtained through responses to the "wallet question", we employ a probit model in the first stage and an ordered probit model in the second stage. Results are reported in Table 7.

Table 5. Online networking and face to face interactions: IV estimates using CMP

|  | Model 1 - SNSs | Model 2 - SNSs with regional controls | Model 3 - Chats, forums, etc. | Model 4 - Chats, forums, etc. with regional controls |
|---|---|---|---|---|
| 1st stage: the dependent variables are indicators of online networking | | | | |
| Regional population covered by DSL | 0.0111*** | 0.00276* | 0.0121*** | 0.000984 |
|  | (8.58) | (1.74) | (6.59) | (0.43) |
| Digital divide (regional area not covered by fibre) | 0.00579*** | 0.00348** | 0.00670** | 0.00395* |
|  | (3.62) | (2.15) | (2.84) | (1.67) |
| dsl | 0.249*** | 0.261*** | 0.247*** | 0.267*** |
|  | (8.32) | (8.77) | (5.77) | (6.28) |
| fiber | 0.273*** | 0.303*** | 0.127 | 0.175* |
|  | (3.83) | (4.29) | (1.28) | (1.79) |
| satellite | 0.297*** | 0.307*** | 0.279*** | 0.303*** |
|  | (7.43) | (7.73) | (4.88) | (5.34) |
| 3G | 0.341*** | 0.360*** | 0.224** | 0.255** |
|  | (5.52) | (5.85) | (2.51) | (2.86) |
| USB | 0.183*** | 0.187*** | 0.115** | 0.134** |
|  | (5.38) | (5.53) | (2.26) | (2.63) |
| mobile | 0.305*** | 0.319*** | 0.349*** | 0.369*** |
|  | (4.92) | (5.17) | (4.96) | (5.27) |
| women | -0.184*** | -0.180*** | -0.244*** | -0.238*** |
|  | (-10.91) | (-10.69) | (-10.18) | (-9.93) |
| age | -0.0863*** | -0.0869*** | -0.0978*** | -0.0978*** |
|  | (-14.34) | (-14.48) | (-12.30) | (-12.40) |
| age squared/100 | 0.0578*** | 0.0587*** | 0.0722*** | 0.0728*** |
|  | (8.03) | (8.17) | (7.53) | (7.66) |
| minutes spent on commuting | -0.00116* | -0.00139** | 0.000000634 | -0.000229 |
|  | (-1.75) | (-2.09) | (0.00) | (-0.25) |
| frequency of meeting friends (by region) |  | 0.178** |  | 0.380** |
|  |  | (2.04) |  | (3.03) |
| high education (% by region) |  | 0.0262*** |  | 0.0271** |
|  |  | (4.23) |  | (3.23) |
| real GDP per capita (thousands euro 2005) |  | -0.0170*** |  | -0.0160*** |
|  |  | (-5.97) |  | (-4.05) |
| 2nd stage: the dependent variable is the frequency of meeting with friends | | | | |
| Participation in Social Networking Sites (d) | 0.950*** | 0.959*** |  |  |
|  | (55.58) | (56.25) |  |  |
| Participation in chats, forums and newsgroups (d) |  |  | 1.067*** | 1.078*** |
|  |  |  | (44.95) | (46.09) |
| N | 35201 | 35201 | 17231 | 17231 |
| F_stat | 73.91 | 5.388 | 43.50 | 2.939 |
| J_stat | 6208.7 | 6496.2 | 6210.7 | 6486.8 |
| chi2 | 8997.5 | 9413.1 | 5055.1 | 5398.3 |

*t* statistics in parentheses. Regressions include socio-demographic and year controls: variables are omitted for the sake of brevity and are available upon request to the authors.
d = for discrete change of dummy variable from 0 to 1
\* p < 0.1, ** p < 0.05, *** p < 0.001

Table 6. Online networking and social trust: IV estimates using CMP

|  | Model 1 - SNSs | Model 2 - SNSs with regional controls | Model 3 - Chats, forums, etc. | Model 3 - Chats, forums, etc. with regional controls |
|---|---|---|---|---|
| 1st stage: the dependent variables are indicators of online networking | | | | |
| Regional population covered by DSL | 0.00662*** | 0.00205 | 0.00804*** | 0.00105 |
|  | (4.60) | (1.15) | (3.79) | (0.40) |
| Digital divide (regional area not covered by fibre) | 0.00817*** | 0.00554** | 0.0113*** | 0.00805** |
|  | (4.70) | (3.13) | (4.36) | (3.08) |
| dsl | 0.306*** | 0.317*** | 0.260*** | 0.280*** |
|  | (9.15) | (9.48) | (5.26) | (5.66) |
| fiber | 0.370*** | 0.399*** | 0.156 | 0.207* |
|  | (4.68) | (5.07) | (1.32) | (1.76) |
| satellite | 0.336*** | 0.346*** | 0.267*** | 0.290*** |
|  | (7.59) | (7.84) | (4.08) | (4.43) |
| 3G | 0.438*** | 0.456*** | 0.317** | 0.352*** |
|  | (6.54) | (6.84) | (3.02) | (3.37) |
| USB | 0.222*** | 0.227*** | 0.122** | 0.138** |
|  | (5.85) | (5.98) | (2.10) | (2.36) |
| mobile | 0.428*** | 0.436*** | 0.496*** | 0.516*** |
|  | (6.29) | (6.41) | (6.26) | (6.49) |
| women | -0.114*** | -0.111*** | -0.176*** | -0.170*** |
|  | (-6.08) | (-5.89) | (-6.41) | (-6.20) |
| age | -0.0580*** | -0.0584*** | -0.0674*** | -0.0677*** |
|  | (-8.73) | (-8.77) | (-7.30) | (-7.33) |
| age squared/100 | 0.0215** | 0.0217** | 0.0348** | 0.0349** |
|  | (2.71) | (2.73) | (3.13) | (3.14) |
| minutes spent on commuting | 0.000112 | -0.000241 | 0.00130 | 0.000925 |
|  | (0.15) | (-0.32) | (1.20) | (0.85) |
| frequency of meeting friends (by region) |  | -0.167* |  | 0.00356 |
|  |  | (-1.73) |  | (0.02) |
| high education (% by region) |  | 0.0441*** |  | 0.0451*** |
|  |  | (6.30) |  | (4.66) |
| real GDP per capita (thousands euro 2005) |  | -0.0212*** |  | -0.0203*** |
|  |  | (-6.61) |  | (-4.40) |

2nd stage: the dependent variable is social trust

|  | Model 1 - SNSs | Model 2 - SNSs with regional controls | Model 3 - Chats, forums, etc. | Model 3 - Chats, forums, etc. with regional controls |
|---|---|---|---|---|
| Participation in Social Networking Sites (d) | -0.242*** | -0.256*** | | |
|  | (-11.44) | (-11.96) | | |
| Participation in chats, forums and newsgroups (d) | | | -0.209*** | -0.230*** |
|  | | | (-6.44) | (-6.97) |
| N | 35197 | 35197 | 17225 | 17225 |
| F_stat | 29.51 | 9.849 | 23.13 | 10.55 |
| J_stat | 7067.5 | 6132.6 | 7071.5 | 6096.3 |
| chi2 | 4036.5 | 4102.9 | 1904.7 | 1945.9 |

*t* statistics in parentheses. Regressions include socio-demographic and year controls: variables are omitted for the sake of brevity and are available upon request to the authors.
d = for discrete change of dummy variable from 0 to 1.
* $p < 0.1$, ** $p < 0.05$, *** $p < 0.001$

Table 7. Online networking and social trust measured through the "wallet question": IV estimates using CMP

|  | Model 1 - SNSs | Model 2 - SNSs with regional controls | Model 3 - Chats, forums, etc. | Model 3 - Chats, forums, etc. with regional controls |
|---|---|---|---|---|
| **1st stage: the dependent variables are indicators of online networking** | | | | |
| Regional population covered by DSL | 0.00685*** (4.71) | 0.00238 (1.33) | 0.00833*** (3.90) | 0.00128 (0.49) |
| Digital divide (regional area not covered by fibre) | 0.00853*** (4.87) | 0.00601*** (3.38) | 0.0118*** (4.52) | 0.00877*** (3.33) |
| dsl | 0.306*** (9.08) | 0.317*** (9.40) | 0.255*** (5.14) | 0.275*** (5.53) |
| fiber | 0.361*** (4.57) | 0.388*** (4.94) | 0.143 (1.22) | 0.193* (1.65) |
| satellite | 0.343*** (7.73) | 0.354*** (7.97) | 0.282*** (4.31) | 0.308*** (4.70) |
| 3G | 0.432*** (6.44) | 0.449*** (6.70) | 0.313** (2.96) | 0.345** (3.27) |
| USB | 0.228*** (5.96) | 0.233*** (6.10) | 0.128** (2.18) | 0.145** (2.48) |
| mobile | 0.433*** (6.35) | 0.441*** (6.47) | 0.502*** (6.31) | 0.522*** (6.56) |
| women | -0.122*** (-6.50) | -0.119*** (-6.35) | -0.187*** (-6.82) | -0.183*** (-6.68) |
| age | -0.0602*** (-9.08) | -0.0610*** (-9.19) | -0.0695*** (-7.51) | -0.0703*** (-7.60) |
| age squared/100 | 0.0248** (3.14) | 0.0254** (3.22) | 0.0378*** (3.42) | 0.0386*** (3.48) |
| minutes spent on commuting | 0.0000334 (0.04) | -0.000313 (-0.42) | 0.00127 (1.16) | 0.000894 (0.82) |
| frequency of meeting friends (by region) |  | -0.152 (-1.56) |  | 0.0433 (0.30) |
| high education (% by region) |  | 0.0459*** (6.48) |  | 0.0472*** (4.82) |
| real GDP per capita (thousands euro 2005) |  | -0.0212*** (-6.55) |  | -0.0200*** (-4.29) |
| **2nd stage: the dependent variable is social trust as measured through the "wallet question"** | | | | |
| Participation in Social Networking Sites (d) | -0.228*** (-12.03) | -0.247*** (-12.71) |  |  |
| Participation in chats, forums and newsgroups (d) |  |  | -0.214*** (-7.44) | -0.239*** (-8.13) |
| N | 35168 | 35168 | 17217 | 17217 |
| F_stat | 31.13 | 11.45 | 24.56 | 12.27 |
| J_stat | 1726.4 | 2042.8 | 1730.9 | 2047.4 |
| chi2 | 3988.6 | 4068.6 | 1889.2 | 1936.1 |

*t* statistics in parentheses. Regressions include socio-demographic and year controls: variables are omitted for the sake of brevity and are available upon request to the authors.
d = for discrete change of dummy variable from 0 to 1.
* $p < 0.1$, ** $p < 0.05$, *** $p < 0.001$

The first stage estimations conducted through probit models show that our instruments satisfy the relevance condition, as their coefficients are statistically significant. The F-statistics (reported at the bottom of Tables 5, 6, and 7) which tests the hypothesis that the coefficient of the excluded instruments are all zero in each first-stage estimate are well above the threshold of 10 (suggested by the literature as the rule of thumb criterion of instrument strength).

To statistically test for correlation of our instruments with the error term of the structural equations (4), we ran an over-identifying restriction test: we used a likelihood ratio test to compare the likelihood function of the two-stage estimates with the likelihood function of a specification, which additionally includes the two instruments. Taken together with the tests of joint significance, the non-rejection of the tests of over-identification suggests that our set of instruments is reasonable.

Addressing endogeneity allowed us to obtain more reliable results on the role of online networking. As reported in Tables 5, 6, and 7, we found that online networking diversely affects the two social capital's dimensions we account for. On the one hand, both participation in SNSs and in chats, forums, and newsgroups seem to support sociability by increasing the likelihood of face-to-face encounters. On the other hand, online networking is found to significantly and negatively affect social trust, however it is measured (i.e. through responses to the "Rosenberg question" or to the "wallet question"). Introducing online networking in regressions makes the statistical significance of commuting disappear.

Women show a significantly lower propensity for face-to-face interaction and significantly lower levels of social trust. Both the frequency of meetings with friends and social trust – however measured – are U-shaped with age.

To assess the robustness of our results, we also considered our dependent variables as continuous variables and we re-estimated our models with a linear 2SLS technique employing the same set of instruments. Results of previous regressions are fully confirmed. Coefficients are reported in Tables 9, 10, and 11 in the Appendix.

The first stages of estimates reported in Tables 9, 10, and 11 highlight the role of DSL and mobile phones in individual access to online networking. The individual-level availability of fibre, which is the fastest way to connect to the Internet, does not significantly influence participation in chats, forums, and newsgroups. Women show a significantly lower propensity for participation in networks like Facebook and Twitter and in chats, forums, and newsgroups. The propensity for participation in social networking sites and in chats, forums, and newsgroups significantly decreases with age.

To compare relative magnitudes of the effects of the independent variables, we computed their marginal effects, which are reported in Table 11. The table also reports the predicted probabilities of meeting friends with a certain frequency (never, less than four times per year, and at least once per week) and of reporting trust in unknown others (as measured through the Rosenberg and the wallet question).

Those who use social networks have a probability of 74% of meeting their friends at least once per week. Facebook and Twitter users, however, show approximately a probability of 28% of thinking that most people cannot be trusted. Participation in SNSs also entails an approximately 87% probability of responding that strangers are 'not very' or 'not at all' likely to return a lost wallet. Estimates are similar regarding the use of chats, forums, and newsgroups.

Marginal effects suggest that as an individual begins using Facebook (or another SNS), the probability of meeting friends frequently (at least once per week) rises by 24%, the probability of thinking that others can be trusted decreases by 8%, and the probability of thinking that a stranger would return a lost wallet decreases by 8%. As an individual begins using chats, forums, and newsgroups, the probability of frequently meeting friends rises by 27%, the probability of thinking that others can be trusted decreases and that a stranger would return a lost wallet decreases by 7%.

Table 8: predicted probabilities and marginal effects

| | Predicted probabilities | | | | Marginal effects | | |
|---|---|---|---|---|---|---|---|
| Frequency of meeting friends | | | | | | | |
| | Never | Less than 4 times per year | At least once per week | | Never | less than 4 times a year | at least once a week |
| SNSs | .028*** | 0.226*** | 0.745*** | SNSs | -0.048*** | -0.187*** | 0.236*** |
| Chats, etc. | 0.028*** | 0.230*** | 0.740*** | Chat, etc. | -.057*** | -.215*** | .272*** |
| Social trust (wallet question) * | | | | | | | |
| | Not much likely or not likely at all | Fairly likely | Very likely | | Not much likely or not likely at all | Fairly likely | Very likely |
| SNSs | 0.869*** | 0.112*** | 0.018*** | SNSs | 0.039*** | -0.031*** | -0.008*** |
| Chat, etc. | 0.866*** | 0.115*** | 0.018*** | Chat, etc. | 0.030*** | -0.023*** | -0.006*** |
| Social trust | | | | | | | |
| | Others can be trusted | | | | Others can be trusted | | |
| SNSs | 0.276*** | | | SNSs | -0.08*** | | |
| Chat, etc. | 0.276*** | | | Chat, etc. | -0.07*** | | |

* "In the city or area where you live, imagine you lost your wallet holding money and your identification or address and it was found by someone else. How likely do you think your wallet would be returned to you if it were found by a stranger?"

Figures 1, 2, and 3 illustrate how the predicted probabilities of never meeting friends, of meeting friends less than four times per year, or at least once per week, vary with age.

Rhombi represent individuals who do not use social networking sites or chats, newsgroups and forums. Figures 1 and 2 show that individuals aged between 40 and 70 who do not use online networking are exposed to a significantly higher risk of being socially isolated. Triangles refer to individuals who only use SNSs, and squares refer to individuals who only use chats, forums, and newsgroups. Circles represent those who use SNSs and chats, forums, and newsgroups. These individuals report a significantly higher probability of meeting friends at least once per week (see Figure 3). Figure 2 shows that the positive effect of online networking on sociability becomes particularly relevant for individuals aged between 30 and 70, when time constraints may be more severe due to work and family obligations.

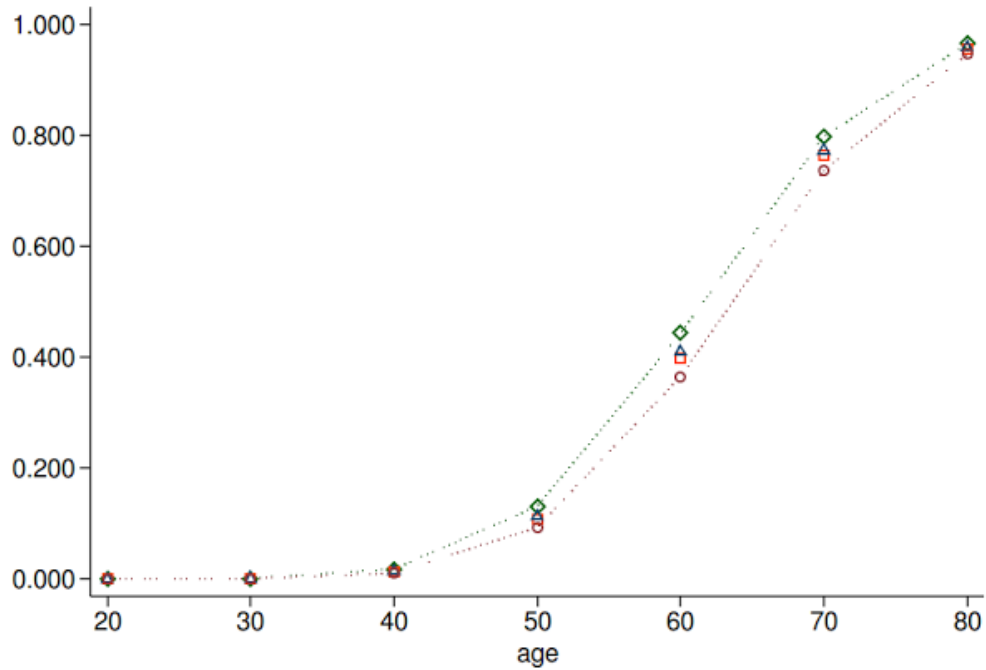

*Figure 1: Predicted probability of never meeting friends by age. Rhombi identify people who do not use SNSs* and chats, newsgroups, and forums*; triangles identify people who attend chats, forums and newsgroups, but not SNSs; squares identify people who attend SNSs, but not chats, forums and newsgroups; circles identify people who use both SNSs and chats, newsgroups, and forums.*

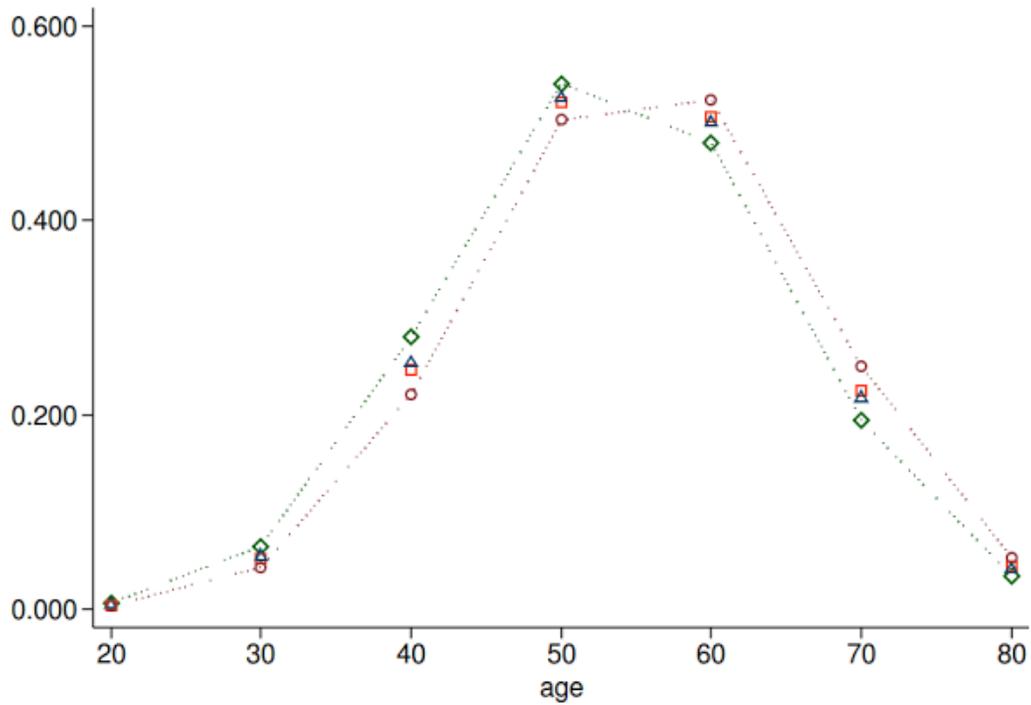

*Figure 2: Predicted probability of meeting friends less than four times per year by age.*

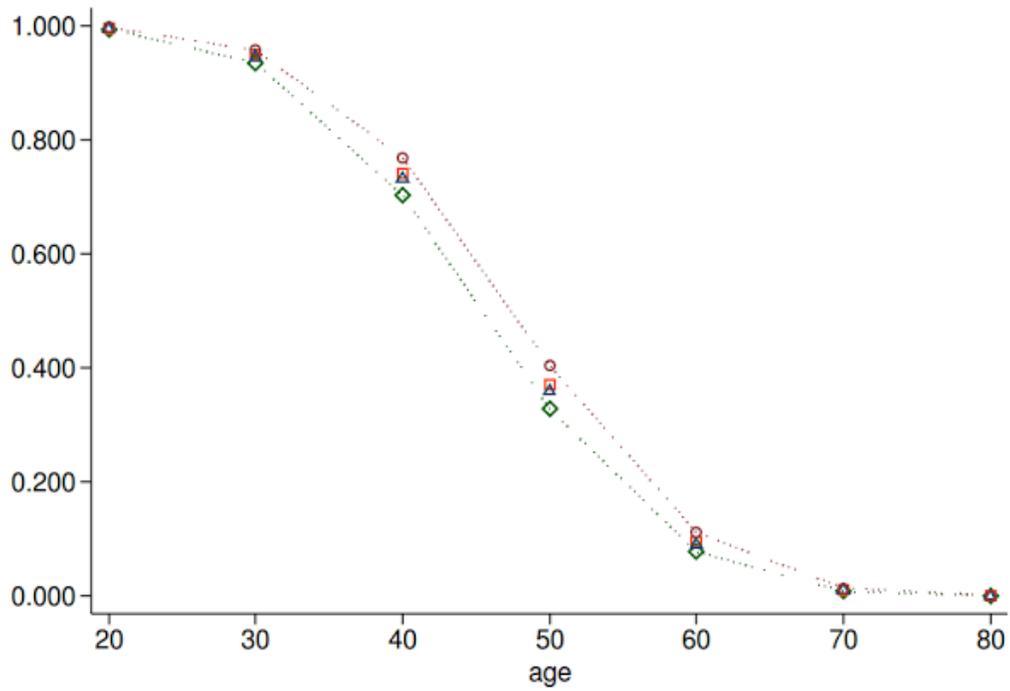

*Figure 3: Predicted probability of meeting friends at least once per week by age.*

# 7. Interpretation of results

The findings reported in Sections VI and VII lead us to argue that, due to the "online networking revolution", Internet use is more likely to support – rather than destroy – sociability and face-to-face interactions. This result contradicts cross-sectional analyses conducted from the late 1990s to the first half of 2000s which argued that time spent browsing the web was positively related to loneliness and negatively related to life-satisfaction (see for example Kraut et al., 1998; Nie and Erbring, 2000; Hamburger and Ben-Artzi, 2003). However, these "pessimistic" findings about the role of the Internet in sociability suffer from two major weaknesses. First, they were conducted before the "social networking revolution" which has made the Internet a fertile environment to nurture social relationships. After the explosion of networks such as Facebook and Twitter, Internet-mediated interaction has become a powerful tool to preserve existing relations and to activate latent ones. Second, most of those studies do not address endogeneity issues, mainly due to the lack of suitable data. As suggested in Section V, people who are already lonely may in fact be more inclined toward Internet use. This unaddressed bias questions the causal relationship between Internet usage and social interactions found in earlier studies. For example, Hamburger and Ben-Artzi (2003) drew on data from a field study on Internet use and feelings of loneliness, extroversion, and neuroticism conducted on 89 participants, to find that lonely people have a tendency to engage in greater Internet usage compared to non-lonely people. Analysing responses from a survey of 277 undergraduate Internet users, Morahan-Martin and Schumacher (2003) showed that "lonely individuals may be drawn online because of the increased potential for companionship, the changed social interaction patterns online, and as a way to modulate negative moods associated with loneliness" (p. 659).

Overall, our results provide support for those more recent empirical studies in the fields of sociology, applied psychology and communication science which found SNSs use to be positively related to face to face interactions and sociability in limited samples of students (see Section III for a review of the literature). Authors of these works claimed that participation in SNSs allows users to preserve and consolidate existing relationships against the threats posed by increasing busyness and mobility (see for example Steinfield et al., 2008). Internet-mediated communication also helps to lower barriers to interaction and encourage self-disclosure. As a result, SNSs users are more likely to activate latent ties which might otherwise remain ephemeral; they encourage interactions with friends or acquaintances of their friends or acquaintances with whom they share interests or relational goals, thereby enabling friendships/relationships which would not otherwise occur. Interaction on Facebook "makes it easier to convert latent ties into weak ties, in that the site provides personal information about others" (Ellison et al., 2007), as the site makes visible one's connections and cultural/political orientation, and enables users to identify individuals with whom they may have some things in common.

Our findings also suggest that online networking displays higher effects on the sociability of individuals aged between 30 and 70. Under 30 individuals are likely to have more opportunities of socialization independently of their participation in online networks through their, for instance, enrolment in secondary or tertiary education programmes. After 30 time constraints are likely to become more severe as the busyness related to family and professional obligations is more likely to increase.

Our result on social trust, on the other hand, conflicts with more optimistic results of psychological and communication studies, which found that SNSs use does not harm social trust in small and biased samples of Facebook users (e.g. college students)

This may be due to the fact that not only online networking allows Internet users to preserve their social ties, but it also favours new contacts with people outside of users' usual reference groups. In face-to-face interactions we usually select a narrow circle of people with whom we discuss values and beliefs (e.g. political and moral issues, such as those related to racism and civil rights). SNSs, by contrast, propose rooms for discussion where selection mechanisms are weak or lacking. Think for example of the Facebook page of a newspaper where a very heterogeneous audience can comment on news and op-ed articles without moderation. In these online discussions, individuals are forced to "meet", and often happen to confront themselves, with a wide variety of points of views. For example, a follower of an anti-abortion movement may actively discuss with those who believe that a woman should have the legal right to abortion, a homophobic individual may be confronted with people who support equality of civil rights (or vice versa), and a Real Madrid's fan will probably discover that Barcelona's supporters are spread all over the world.

Diversity is much more diffused in the global population of Internet users than in their limited reference groups. Empirical studies have shown that, at least in the short run, diversity along ethnic, religious, age, and socio-economic status lines may be a powerful source of frustration and distrust towards unknown others (Subramanian et al., 2002; Alesina and La Ferrara 2006; Christoforou, 2011).

Another source of frustration and distrust could be related to the fact that in the Internet-mediated interaction with strangers, individuals often exhibit a higher propensity for aggressive behaviour than in face-to-face interactions. In public online forums for discussion – such as those offered by Facebook's "public pages" (like those managed by public figures, newspapers, political movements, etc.), "groups", and "communities", and by commenting platforms for online magazines and newspapers (e.g. Disqus, IntenseDebate, Livefyre) – individuals are likely to deal with strangers in a more aggressive and unscrupulous way than they would in a physical meeting. In online environments, unknown strangers basically are "invisible" and their reaction to provocative behaviours may be easily neutralised (for example by simply withdrawing from the conversation, or even by "blocking" them through the network's privacy settings). In addition, online conversations are more vulnerable to incomprehension and misunderstandings. Face to face interactions, by contrast, allow better articulation of one's expressions, gestures, the tone of voice, feelings, opinions, and intentions, but disallow the possibility of easily withdrawing from unpleasant conversations. Thus, physical interactions with people are less aggressive.

In our Italian case study, the anectodal evidence suggests that the practice of hate speech is rapidly spreading in online networks. For example, on October 2013, a boat carrying migrants from Libya to Italy sank off the Italian Southern island of Lampedusa. More than 360 African migrants died. According to Italian police's reports, dozens of survivors were raped and tortured in Libya before starting their journey. A second shipwreck occurred a few days later. The boat was reportedly carrying migrants from Syria and Palestine, and at least 34 individuals were later confirmed dead. A quick scroll through the comments on many newspapers' Facebook pages revealed thousands of racist messages posted on almost every article about the Lampedusa disasters and about the personal stories of the people who died in the shipwrecks. Most messages expressed feelings of pleasure and satisfaction with migrant tragedies, and the wish for more boats to sank.

This example, which recently triggered the public debate on hate speech in Italy, is just a taste of the wide range of prejudiced or violent comments on Facebook pages. The psychological impact of these comments is likely to be detrimental for individuals' trust toward strangers.

Our finding that being a woman significantly raise the probability to loose social trust as a consequence of online interactions can be explained by women's higher likelihood to be targeted by hate speech. According to a 2005 report by the Pew Research Center, women and men have been logging on in equal numbers since 2000, but the vilest and most harsh communications are still disproportionately lobbed at women (Fallows, 2005).

Women are more likely to report being stalked and harassed on the Internet. 72.5 per cent of the 3,787 people who reported harassing incidents from 2000 to 2012 were women, according to the volunteer organization Working to Halt Online Abuse (WHOA). Sometimes, the abuse can get physical: A Pew survey reported that five per cent of women who used the Internet said "something happened online" that led them into "physical danger." And it starts young: teenage girls are significantly more likely to be cyberbullied than boys. Journalist Amanda Hess reports in the *Pacific Standard* that: "Just appearing as a woman online, it seems, can be enough to inspire abuse. In 2006, researchers from the University of Maryland set up a bunch of fake online accounts and then dispatched them into chat rooms. Accounts with feminine usernames incurred an average of 100 sexually explicit or threatening messages a day. Masculine names received 3.7".

In 2012 activist groups Women, Action, and the Media (WAM) and the Everyday Sexism Project started the #FBRape campaign with a call for Facebook users to contact companies whose ads were appearing on pages beside the violent and misogynist content and call for them to withdraw their advertising from the site. As reported by Ryan Lenora Brown in *The Christian Science Monitor*, at issue were not just the violent images themselves, but also the fact that Facebook was failing to delete them when users flagged the photos as hate speech.

The contradiction between our findings on social trust and results from previous literature should also be interpreted in relation to whom respondents have in mind when they answer to the question: "Generally speaking, do you think that most people can be trusted?" In psychological studies based on small groups of undergraduate students, the "radius of trust" (Fukuyama, 1999) may well be limited to the respondents' small circles of fellow students and friends. In our nationally and regionally representative sample of the Italian population the radius of trust is likely to be much more extended. Previous studies have shown that the further people move from their immediate circle of friends, colleagues, and neighbour, the more sceptical they are (Delhey et al., 2011; Welch et al., 2007). As stated by Delhey et al. (2011), "Differences in trust levels can be interpreted sensibly only when trust radiuses are similar" (p. 789). Unfortunately MHS data do not allow us to control the radius of trust. However, this argument suggests caution in the comparison of our results with previous findings in psychology and communication science literature and urges researchers to use larger and more representative samples for investigating the overall role of networking on values, beliefs, and pro-social behaviours.

## 8. Conclusions

Will Internet usage accelerate the decline in social participation documented by empirical studies in the "social networking era"? Or does it offer a way to support social relationships against the threats posed by the disruption of ties and the weakening of community life? How does online networking affect trust? In this paper we empirically analysed how participation in networking sites like Facebook and Twitter affects two main dimensions of social capital, i.e., frequency of face to face interactions and social trust. The empirical analysis used a pooled cross-section of data including 50,000 observations from the last two waves (2010 and 2011) of the Istat Multipurpose Survey on Households (MHS). This survey contains detailed information on individual propensity for Internet-mediated interaction through participation in social networking sites and in chats, newsgroups, and forums. The dataset also includes information on several aspects of individual well-being and on a number of social capital's dimensions such as relationships with friends and acquaintances, shared values and beliefs, and social trust. Given the cross-sectional nature of the sample, our identification strategy basically relies on the use of two indicators of technological infrastructures – which may be considered as an exogenous aspect of the diffusion of high-speed connections to the Internet across Italian regions – as instruments for online networking.

Our findings suggest that the online networking revolution is allowing the Internet to support – rather than destroy – sociability and face-to-face interactions. Social networking seems to offer a powerful tool to protect social relationships against the threats posed by increasing busyness and mobility. This result is consistent with previous analyses conducted on small samples in the fields of social psychology and communication science.

This result suggests that the digital divide is likely to become an increasingly important factor of social exclusion that may significantly exacerbate inequalities in well-being and capabilities.

The result on social trust, however, contrasts with common findings in the aforementioned fields of studies. We suggest that the decline in trust may be interpreted as an individual reaction to diversity, which has been found to be a major source of frustration and distrust by empirical studies in economics. The conflicting directions of online networking's effects on social networks and social trust also suggest that Internet usage may be reinforcing the distinction between in-group and out-group relationships, as far as it seems to help individuals to further strengthen their social relationships and to lower their trust in unknown others.

It is worth noting that our result on social trust seems to be in line with studies finding that the diffusion of broadband Internet has initially produced disenchantment among web surfers, in turn causing a fall in electoral participation (Falck et al. 2012; Campante et al. 2013).

Our analysis does not account for all relevant dimensions of social capital which should be further investigated in a follow up of this study. Furthermore, we do not claim to have solved endogeneity issues. Rather, the cross-sectional nature of our study definitely suggests caution in the interpretation of results as dictated by causal relationships. In addition, results about social trust definitely require more analysis and interpretation. But our study represents the first attempt to investigate the role of online networking in social interactions in a large and representative sample, and it provides evidence that participation in SNSs does not necessarily favour social isolation.

# Appendix

Table 9. Online networking and face to face interactions: IV estimates

| | Model 1 - SNSs | | Model 2 - SNSs with regional controls | | Model 3 - Chats, forums, etc. | | Model 4 - Chats, forums, etc. with regional controls | |
|---|---|---|---|---|---|---|---|---|
| | 1st stage | 2nd stage | 1st stage | 2nd stage | 1st stage | 2nd stage | 1st stage | 2nd stage |
| Regional population covered by DSL | 0.00582*** (4.13) | | 0.00247 (1.41) | | 0.00747*** (4.40) | | 0.00111 (0.43) | |
| Digital divide (regional area not covered by fibre) | 0.00832*** (4.86) | | 0.00618*** (3.54) | | 0.00989*** (4.68) | | 0.00865*** (3.35) | |
| dsl | 0.306*** (9.22) | -0.263*** (-3.58) | 0.316*** (9.50) | 0.0550 (0.76) | 0.159*** (4.04) | -0.0876 (-1.31) | 0.286*** (5.84) | 0.209** (2.35) |
| fiber | 0.355*** (4.61) | -0.436*** (-3.75) | 0.378*** (4.91) | -0.0152 (-0.14) | 0.0391 (0.41) | -0.193* (-1.78) | 0.195* (1.71) | 0.0650 (0.55) |
| satellite | 0.350*** (8.00) | -0.286** (-3.25) | 0.361*** (8.25) | 0.0803 (0.94) | 0.118** (2.18) | -0.0958 (-1.16) | 0.322*** (4.99) | 0.249** (2.32) |
| 3G | 0.412*** (6.23) | -0.393*** (-3.31) | 0.427*** (6.46) | 0.0476 (0.42) | 0.157* (1.91) | -0.217* (-1.81) | 0.353*** (3.38) | 0.150 (1.09) |
| USB | 0.226*** (6.00) | -0.185** (-2.89) | 0.231*** (6.13) | 0.0320 (0.56) | 0.0987** (2.17) | -0.0693 (-1.16) | 0.154** (2.67) | 0.0862 (1.30) |
| mobile | 0.430*** (6.39) | -0.488*** (-3.98) | 0.436*** (6.47) | -0.0312 (-0.28) | 0.401*** (5.70) | -0.346** (-2.83) | 0.512*** (6.52) | 0.199 (1.24) |
| women | -0.123*** (-6.62) | -0.0770** (-2.19) | -0.122*** (-6.52) | -0.188*** (-5.97) | -0.0344 (-1.52) | -0.106** (-2.40) | -0.184*** (-6.74) | -0.268*** (-4.79) |
| age | -0.0609*** (-9.38) | -0.0199 (-1.20) | -0.0614*** (-9.45) | -0.0888*** (-5.40) | -0.0853*** (-11.08) | -0.0454** (-2.32) | -0.0695*** (-7.66) | -0.134*** (-5.12) |
| age squared/100 | 0.0262*** (3.39) | 0.0478*** (4.29) | 0.0266*** (3.43) | 0.0814*** (8.19) | 0.0721*** (8.30) | 0.0528*** (3.33) | 0.0381*** (3.50) | 0.115*** (5.72) |
| minutes spent on commuting | -0.000219 (-0.30) | -0.00237** (-2.82) | -0.000532 (-0.73) | -0.00222*** (-3.81) | 0.00128 (1.39) | -0.00254** (-2.60) | 0.000493 (0.46) | -0.00148 (-1.46) |
| frequency of meeting friends (by region) | | | -0.127 (-1.33) | 0.662*** (7.99) | | | 0.0324 (0.23) | 0.699*** (5.48) |
| high education (% by region) | | | 0.0396*** (5.74) | -0.00709 (-0.65) | | | 0.0390*** (4.10) | 0.0149 (1.03) |
| real GDP per capita (thousands euro 2005) | | | -0.0167*** (-5.35) | -0.00996** (-2.06) | | | -0.0169*** (-3.77) | -0.0183** (-2.83) |
| Participation in Social Networking Sites (d) | | 2.701*** (4.23) | | -0.245 (-0.37) | | | | |
| Participation in chats, forums and newsgroups (d) | | | | | | 1.564** (2.64) | | -1.472* (-1.76) |
| Constant | -0.999* (-1.80) | 5.196*** (7.87) | 0.105 (0.13) | 4.646*** (5.12) | -0.126 (-0.26) | 5.737*** (5.75) | -0.0797 (-0.07) | 5.426*** (3.72) |
| N | 22204 | 22148 | 22204 | 22148 | 21050 | 10765 | 10790 | 10765 |
| F_stat | 28.02 | | 12.54 | | 28.04 | | 12.55 | |
| J_stat | | | | | | | | |
| chi2 | 4106.8 | 3610.7 | 4138.5 | 8408.1 | 7191.3 | 2813.1 | 1950.9 | 2902.9 |

*t* statistics in parentheses.
a: The first stage has indicators of online networking as dependent variables.
b: In the second stage, dependent variables are indicators of social capital.
Regressions include socio-demographic and year controls: variables are omitted for the sake of brevity and are available upon request to the authors.
d = for discrete change of dummy variable from 0 to 1
t values in brackets
* $p < 0.1$, ** $p < 0.05$, *** $p < 0.001$

Table 10. Online networking and social trust: IV estimates

| | Model 1 - SNSs | | Model 2 - SNSs with regional controls | | Model 3 - Chats, forums, etc. | | Model 4 - Chats, forums, etc. with regional controls | |
|---|---|---|---|---|---|---|---|---|
| | 1st stage | 2nd stage | 1st stage | 2nd stage | 1st stage | 2nd stage | 1st stage | 2nd stage |
| Regional population covered by DSL | 0.00582*** | | 0.00247 | | 0.00747*** | | 0.00111 | |
| | (4.13) | | (1.41) | | (4.40) | | (0.43) | |
| Digital divide (regional area not covered by fibre) | 0.00832*** | | 0.00618*** | | 0.00989*** | | 0.00865*** | |
| | (4.86) | | (3.54) | | (4.68) | | (3.35) | |
| dsl | 0.306*** | 0.125*** | 0.316*** | 0.0474 | 0.159*** | 0.0552** | 0.286*** | 0.0106 |
| | (9.22) | (4.10) | (9.50) | (1.50) | (4.04) | (2.11) | (5.84) | (0.36) |
| fiber | 0.355*** | 0.164*** | 0.378*** | 0.0576 | 0.0391 | 0.0438 | 0.195* | -0.00115 |
| | (4.61) | (3.33) | (4.91) | (1.24) | (0.41) | (0.95) | (1.71) | (-0.03) |
| satellite | 0.350*** | 0.167*** | 0.361*** | 0.0771** | 0.118** | 0.0855** | 0.322*** | 0.0741** |
| | (8.00) | (4.52) | (8.25) | (2.06) | (2.18) | (2.81) | (4.99) | (2.08) |
| 3G | 0.412*** | 0.181*** | 0.427*** | 0.0725 | 0.157* | 0.0768* | 0.353*** | 0.0670 |
| | (6.23) | (3.69) | (6.46) | (1.53) | (1.91) | (1.96) | (3.38) | (1.42) |
| USB | 0.226*** | 0.0986*** | 0.231*** | 0.0462* | 0.0987** | 0.0325 | 0.154** | 0.0281 |
| | (6.00) | (3.74) | (6.13) | (1.88) | (2.17) | (1.32) | (2.67) | (1.27) |
| mobile | 0.430*** | 0.184*** | 0.436*** | 0.0761 | 0.401*** | 0.226** | 0.512*** | 0.0480 |
| | (6.39) | (3.61) | (6.47) | (1.58) | (5.70) | (3.25) | (6.52) | (0.91) |
| women | -0.123*** | -0.0688*** | -0.122*** | -0.0416** | -0.0344 | -0.0397*** | -0.184*** | -0.0297 |
| | (-6.62) | (-4.64) | (-6.52) | (-2.98) | (-1.52) | (-3.45) | (-6.74) | (-1.59) |
| age | -0.0609*** | -0.0219** | -0.0614*** | -0.00502 | -0.0853*** | -0.0278** | -0.0695*** | 0.00295 |
| | (-9.38) | (-3.13) | (-9.45) | (-0.69) | (-11.08) | (-3.09) | (-7.66) | (0.34) |
| age squared/100 | 0.0262*** | 0.0153** | 0.0266*** | 0.00703 | 0.0721*** | 0.0282*** | 0.0381*** | 0.00252 |
| | (3.39) | (3.27) | (3.43) | (1.63) | (8.30) | (3.46) | (3.50) | (0.37) |
| minutes spent on commuting | -0.000219 | -0.000549 | -0.000532 | -0.000434* | 0.00128 | 0.000428 | 0.000493 | -0.000172 |
| | (-0.30) | (-1.55) | (-0.73) | (-1.69) | (1.39) | (0.96) | (0.46) | (-0.48) |
| frequency of meeting friends (by region) | | | -0.127 | 0.00109 | | | 0.0324 | 0.0632 |
| | | | (-1.33) | (0.03) | | | (0.23) | (1.36) |
| high education (% by region) | | | 0.0396*** | -0.0109** | | | 0.0390*** | -0.0173*** |
| | | | (5.74) | (-2.28) | | | (4.10) | (-3.65) |
| real GDP per capita (thousands euro 2005) | | | -0.0167*** | 0.00868*** | | | -0.0169*** | 0.0121*** |
| | | | (-5.35) | (3.98) | | | (-3.77) | (5.47) |
| Participation in Social Networking Sites (d) | | -1.056*** | | -0.339 | | | | |
| | | (-3.91) | | (-1.16) | | | | |
| Participation in chats, forums and newsgroups (d) | | | | | | -1.517*** | | -0.0887 |
| | | | | | | (-3.82) | | (-0.33) |
| Constant | -0.999* | 0.929*** | 0.105 | 0.236 | -0.126 | 1.480*** | -0.0797 | -0.0278 |
| | (-1.80) | (3.69) | (0.13) | (0.61) | (-0.26) | (4.16) | (-0.07) | (-0.06) |
| N | 22204 | 22074 | 22204 | 22074 | 21050 | 20944 | 10790 | 10738 |
| F_stat | 28.02 | | 12.54 | | 28.04 | | 12.55 | |
| J_stat | | | | | | | | |
| chi2 | 4106.8 | 366.5 | 4138.5 | 866.7 | 7191.3 | 286.4 | 1950.9 | 551.4 |

*t* statistics in parentheses.
a: The first stage has indicators of online networking as dependent variables.
b: In the second stage, dependent variables are indicators of social capital.
Regressions include socio-demographic and year controls: variables are omitted for the sake of brevity and are available upon request to the authors.
d = for discrete change of dummy variable from 0 to 1
t values in brackets
* p < 0.1, ** p < 0.05, *** p < 0.001

Table 11. Online networking and social trust (measured through the "wallet question": IV estimates

|  | Model 1 - SNSs | | Model 2 - SNSs with regional controls | | Model 3 - Chats, forums, etc. | | Model 4 - Chats, forums, etc. with regional controls | |
|---|---|---|---|---|---|---|---|---|
|  | 1st stage | 2nd stage | 1st stage | 2nd stage | 1st stage | 2nd stage | 1st stage | 2nd stage |
| Regional population covered by DSL | 0.00582*** (4.13) |  | 0.00247 (1.41) |  | 0.00747*** (4.40) |  | 0.00111 (0.43) |  |
| Digital divide (regional area not covered by fibre) | 0.00832*** (4.86) |  | 0.00618*** (3.54) |  | 0.00989*** (4.68) |  | 0.00865*** (3.35) |  |
| dsl | 0.306*** (9.22) | 0.342*** (4.88) | 0.316*** (9.50) | 0.211** (2.99) | 0.159*** (4.04) | 0.216*** (3.47) | 0.286*** (5.84) | 0.178** (2.73) |
| fiber | 0.355*** (4.61) | 0.482*** (4.22) | 0.378*** (4.91) | 0.304** (2.95) | 0.0391 (0.41) | 0.167 (1.48) | 0.195* (1.71) | 0.162* (1.75) |
| satellite | 0.350*** (8.00) | 0.418*** (4.88) | 0.361*** (8.25) | 0.265** (3.13) | 0.118** (2.18) | 0.220** (2.99) | 0.322*** (4.99) | 0.214** (2.68) |
| 3G | 0.412*** (6.23) | 0.510*** (4.46) | 0.427*** (6.46) | 0.331** (3.09) | 0.157* (1.91) | 0.228** (2.44) | 0.353*** (3.38) | 0.298** (2.80) |
| USB | 0.226*** (6.00) | 0.224*** (3.67) | 0.231*** (6.13) | 0.135** (2.41) | 0.0987** (2.17) | 0.110* (1.89) | 0.154** (2.67) | 0.0756 (1.56) |
| mobile | 0.430*** (6.39) | 0.472*** (3.99) | 0.436*** (6.47) | 0.288** (2.63) | 0.401*** (5.70) | 0.637*** (3.68) | 0.512*** (6.52) | 0.302** (2.47) |
| women | -0.123*** (-6.62) | -0.111** (-3.28) | -0.122*** (-6.52) | -0.0669** (-2.16) | -0.0344 (-1.52) | -0.0366 (-1.27) | -0.184*** (-6.74) | -0.0660 (-1.55) |
| age | -0.0609*** (-9.38) | -0.0569*** (-3.56) | -0.0614*** (-9.45) | -0.0286* (-1.78) | -0.0853*** (-11.08) | -0.0742*** (-3.35) | -0.0695*** (-7.66) | -0.0287 (-1.45) |
| age squared/100 | 0.0262*** (3.39) | 0.0262** (2.42) | 0.0266*** (3.43) | 0.0121 (1.26) | 0.0721*** (8.30) | 0.0642** (3.22) | 0.0381*** (3.50) | 0.0195 (1.28) |
| minutes spent on commuting | -0.000219 (-0.30) | -0.000886 (-1.06) | -0.000532 (-0.73) | -0.000750 (-1.25) | 0.00128 (1.39) | 0.000550 (0.50) | 0.000493 (0.46) | -0.000283 (-0.34) |
| frequency of meeting friends (by region) |  |  | -0.127 (-1.33) | -0.150* (-1.79) |  |  | 0.0324 (0.23) | -0.0961 (-0.94) |
| high education (% by region) |  |  | 0.0396*** (5.74) | -0.0174 (-1.64) |  |  | 0.0390*** (4.10) | -0.0262** (-2.33) |
| real GDP per capita (thousands euro 2005) |  |  | -0.0167*** (-5.35) | 0.0111** (2.34) |  |  | -0.0169*** (-3.77) | 0.0125** (2.51) |
| Participation in Social Networking Sites (d) |  | -3.086*** (-5.07) |  | -1.893** (-2.95) |  |  |  |  |
| Participation in chats, forums and newsgroups (d) |  |  |  |  |  | -4.463*** (-4.60) |  | -1.684** (-2.69) |
| Constant | -0.999* (-1.80) | 3.872*** (6.70) | 0.105 (0.13) | 3.550*** (4.14) | -0.126 (-0.26) | 4.731*** (5.23) | -0.0797 (-0.07) | 3.250** (3.22) |
| N | 22204 | 22081 | 22204 | 22081 | 21050 | 20922 | 10790 | 10729 |
| F_stat | 28.02 |  | 12.54 |  | 28.04 |  | 12.55 |  |
| J_stat |  |  |  |  |  |  |  |  |
| chi2 | 4106.8 | 319.6 | 4138.5 | 937.0 | 7191.3 | 98.58 | 1950.9 | 322.8 |

$t$ statistics in parentheses.
a: The first stage has indicators of online networking as dependent variables.
b: In the second stage, dependent variables are indicators of social capital.
Regressions include socio-demographic and year controls: variables are omitted for the sake of brevity and are available upon request to the authors.
d = for discrete change of dummy variable from 0 to 1
t values in brackets
* $p < 0.1$, ** $p < 0.05$, *** $p < 0.001$

Figure 4: Percentage of the population covered by broadband in Italy and topographic map of Italy.

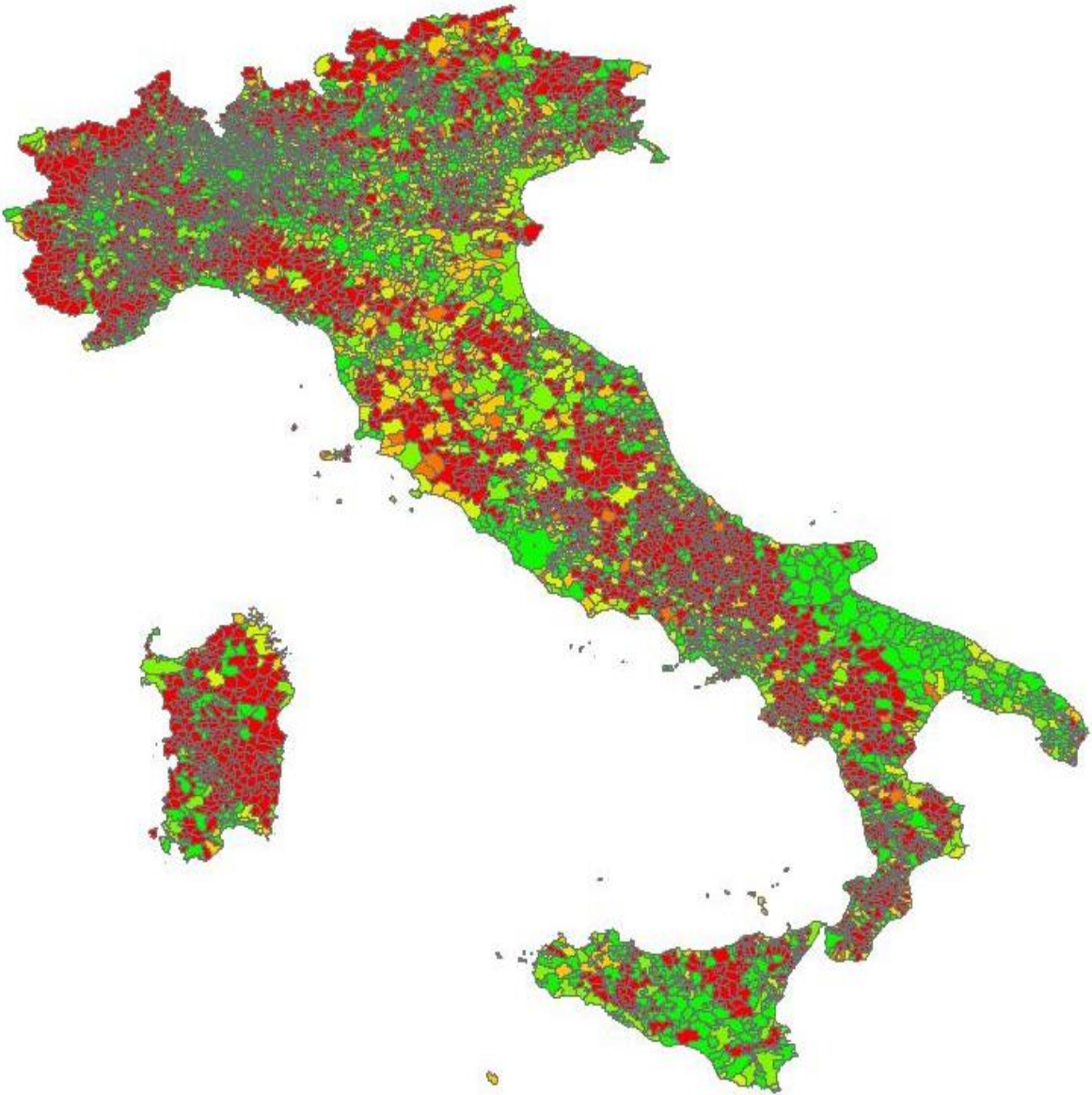

Source: Between (2006), p. 17. Darker areas are those with the worst coverage. Green areas have the best coverage.

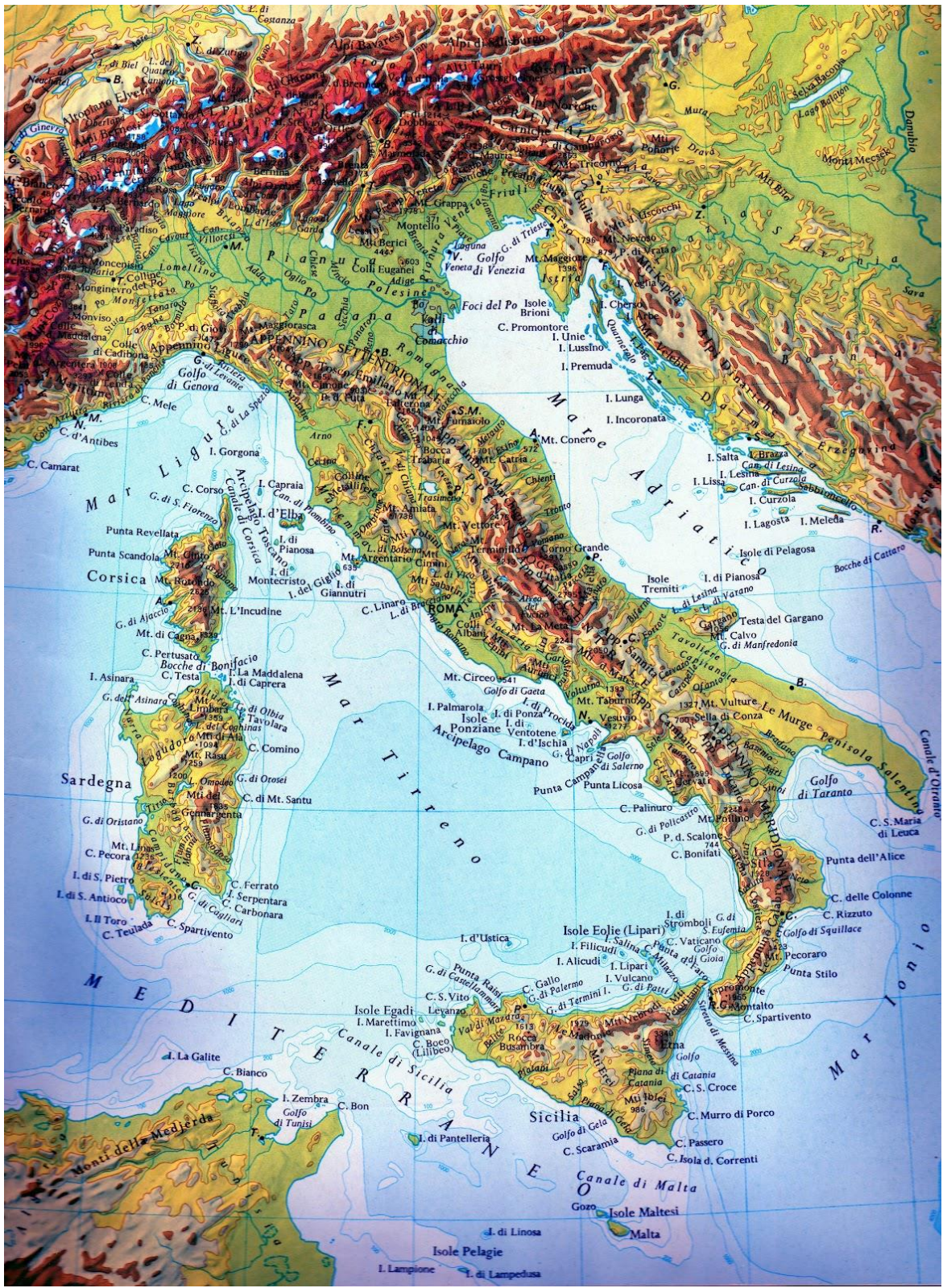

# Endnotes

[1] Both the structural and cognitive dimensions include several sub-dimensions whose relationships with outcome variable s in turn vary according to the context and the effect of other individual and local potentially influential factors (Sabatini, 2008; Degli Antoni and Sacconi, 2009; 2011; Yamamura, 2011a). In addition, structural and cognitive dimensions influence each other. Drawing on Granovetter's (1992) discussion on structural and relational embeddedness, other authors prefer to classify the multiple facets of social capital into three clusters comprising the structural, the relational, and the cognitive dimensions of the concept (see for example Nahapiet and Goshal, 1998). In this three-dimensional classification, structural embeddedness refers to the characteristics of the social system as a whole, and the expression "structural networks' is used to describe impersonal relations among people or groups. By contrast, relational embeddedness refers to personal relations that individuals have developed through a history of interactions (Granovetter, 1992). Cognitive social capital, on the other hand, refers to "those resources providing shared representations, interpretations, and systems of meaning among parties" (Nahapiet and Goshal, 1998: 244).

[2] Following Knack and Keefer (1997), the literature generally distinguishes two types of formal organisations, labelled "Olsonian" and "Putnam-esque" associations. Olson groups are those associations with redistributive goals that lobby for the protection of their members' interests, possibly against the interests of other groups (Olson 1965, 1982). Examples of this type of organisation are professional and entrepreneurial associations, trade unions and associations for the protection of consumers' rights. Putnam groups are those associations least likely to act as "distributional coalitions but which involve social interactions that can build trust and cooperative habits" (Knack & Kefeer, 1997, p. 1273). Examples of this type of organization are cultural circles, sport clubs, youth associations (e.g. scouts) and religious organisations.

[3] Despite the many studies documenting the decline in social participation, the overall evidence still seems to be non-conclusive. A number of empirical studies have found conflicting results on the trends of different indicators of social capital, and the *Bowling Alone* thesis has been variously characterised as plainly wrong, pessimistic or traditional (Stolle and Hooghe, 2005). Worms (2000) and Van Ingen and Dekker (2011) argue that the decline in associational participation may be related to a process of "informalisation" of social activities. In his cross-country analysis of social capital trends, Sarracino (2010) finds that in most Western European countries, several measures of connectedness experienced a growth over the period 1980–2000.

[4] There is different evidence on the social effects of commuting outside of the United States. In countries where cities are, on average, significantly smaller than in the U.S., Putnam's thesis seems not to be supported. A Swiss study by Viry et al. (2009) concludes that while commuting decreases the availability of emotionally bonding social capital in the form of supportive strong ties, it could provide increased opportunities for developing bridging social capital and weak ties.

[5] It is worth noting that part of the literature does not agree with the above reported claims about the beneficial effects of Internet-mediated interaction on social capital. Some studies warn that, beyond a certain threshold, the development of human relationships by the exclusive means of online interactions may prevent users from enjoying those emotional benefits normally associated with face-to-face interactions (see, for example, Lee et al., 2011). Kross et al. (2013) use a sample of 82 people recruited through flyers posted around Ann Arbor, Michigan to analyse the effect of Facebook use on subjective well-being. Five times per day, participants were text-messaged the url of an online survey. The authors find that Facebook use predicts a negative shift in life-satisfaction in their sample. A survey of the literature accurately describing the different positions on the role of Internet-mediated interaction in the accumulation of social capital is included in Antoci et al. (2013a).

[6] Other possible responses were 2 = never, 3 = a few times per year, 4 = less than four times per month, 5 = once per week, 6 = more than once per week.

[7] Possible work status were employed, unemployed looking for a job, first job seeker, household, student, disabled worker, retired worker, other.

[8] According to data provided by Facebook Advertising Platform, in January 2008 Facebook had 216,000 subscribers in Italy. As of October 2013, the network officially reports having 26,000,000 subscribers. Some data are publicly retrievable on the website of the Italian Observatory on Facebook run by Vincenzo Cosenza at the url: http://vincos.it/osservatorio-facebook/.

[9] IV estimates were calculated through Roodman's (2009) Stata module to implement conditional mixed process (cmp) estimator.